\documentclass[fleqn,usenatbib]{mnras}
\usepackage{float}
\usepackage[T1]{fontenc}
\DeclareRobustCommand{\VAN}[3]{#2}
\let\VANthebibliography\thebibliography
\def\thebibliography{\DeclareRobustCommand{\VAN}[3]{##3}\VANthebibliography}
\usepackage{amsmath}	
\usepackage{amssymb}	
\usepackage{graphicx}
\usepackage{graphics}

\title[The blazar PKS 0735+178]{Multi-wavelength spectral modelling of the candidate neutrino blazar
PKS 0735+178}

\author[Athira M. Bharathan et al.]{
Athira M Bharathan$^{1}$\thanks{E-mail: athira.bharathan@res.christuniversity.in}
C. S. Stalin$^{2}$,
S. Sahayanathan$^{3,4}$,
Subir Bhattacharyya$^{3,4}$,
Blesson Mathew$^{1}$
\\
$^{1}$Department of Physics and Electronics, CHRIST (Deemed to be University), Bangalore, India\\
$^{2}$Indian Institute of Astrophysics, Block II, Koramangala, Bangalore 560 034, India\\
$^{3}$Astrophysical Sciences Division, Bhabha Atomic Research Centre, Mumbai - 400085, India \\
$^{4}$Homi Bhabha National Institute, Mumbai 400094, India}

\date{Accepted XXX. Received YYY; in original form ZZZ}

\pubyear{2015}

\begin{document}
\label{firstpage}
\pagerange{\pageref{firstpage}--\pageref{lastpage}}
\maketitle

\begin{abstract}
The BL Lac object PKS 0735+178 was in its historic $\gamma$-ray brightness
state during December 2021. This period also coincides with the detection of a neutrino event
IC211208A, which was localized close to the vicinity of PKS 0735+178. We carried
out detailed $\gamma$-ray timing and spectral analysis of the source in 
three epochs (a) quiescent state ($E_{1}$), (b) moderate activity state ($E_{2}$) and
(c) high activity state ($E_{3}$) coincident with the epoch of neutrino detection. 
During the epoch of neutrino detection ($E_{3}$), we found the largest variability amplitude 
of 95\%.  The $\gamma$-ray spectra corresponding to these three epochs 
are well fit by the power law model and the source is found to show spectral variations 
with a softer when brighter trend. 
In the epoch $E_{3}$, we found the shortest flux
doubling/halving time of 5.75 hrs. Even though the spectral energy distribution in the moderate activity state and in the high activity state could be modeled by the one-zone leptonic emission model, the spectral energy distribution in the quiescent state required an additional component of radiation over and above the leptonic component. Here we show that a photo-meson process was needed to explain the excess $\gamma$-ray emission in the hundreds of GeV which could not be accounted for by the synchrotron self-Compton process.

\end{abstract}




\section{Introduction}

Blazars, among the persistent luminous objects in the Universe with 
luminosity ranging from 10$^{42}$ $-$ 10$^{48}$ erg/sec are a class of active 
galactic nuclei (AGN). They are believed to be powered by the accretion of matter 
onto supermassive black hole with masses of the order of  10$^6$ $-$ 10$^{10}$ 
M$_{\odot}$ located at the centre of galaxies \citep{1969Natur.223..690L,1973A&A....24..337S}. 
Their  radiation is 
dominated by non-thermal processes from relativistic jets
that extend from scales of pc to kpc and are oriented close to 
the line of sight to the observer. They show rapid and large 
amplitude flux variations across the electromagnetic spectrum on time scales 
ranging from minutes to years \citep{1995ARA&A..33..163W,1997ARA&A..35..445U}. 
In addition to flux variations, the optical emission from blazars is highly polarized
\citep{1980ARA&A..18..321A} with significant polarization variations 
\citep{2005A&A...442...97A,2017ApJ...835..275R,2022MNRAS.510.1809P}. 
They are also believed to be cosmic accelerators and could be the sources of
astrophysical neutrinos \citep{1993PhRvD..48.2408M,2020ATel14238....1P,2018Sci...361.1378I,2021Univ....7..492G,2022Univ....8..513B,2023arXiv230511263B}.

Blazars are conventionally divided into
two sub-classes namely  flat spectrum radio quasars (FSRQs) and 
BL Lacertae type objects (BL Lacs) based on the optical spectral 
properties. FSRQs show broad emission lines in their optical spectra
while BL Lacs have either featureless optical spectra or spectra
with weak emission lines with equivalent widths lesser than 5 \AA.
According to \cite{2011MNRAS.414.2674G}, a more physical distinction
between FSRQs and BL Lacs could be based on the luminosity of the 
broad line region (BLR) with FSRQs having L$_{BLR}$/L$_{Edd}$ $>$ 
5 $\times$ 10$^{-4}$. Here, L$_{Edd}$ is the Eddington luminosity 
given by L$_{Edd}$ = 1.3 $\times$ 10$^{38}$ (M/M$_{\odot}$) erg s$^{-1}$.
The broadband spectral energy distribution (SED) of blazars 
displays two prominent peaks.  The low energy component
that peaks between the infrared and X-ray bands 
is well understood as the synchrotron emission from a relativistic population 
of electrons in the blazar jet. The origin of the high energy component, peaking at  
the MeV – GeV energy range \citep{1998MNRAS.299..433F}, is 
highly debated with models that promote a radiation scenario involving either leptons or hadrons or combination of both \citep{2013ApJ...768...54B,2020Galax...8...72C}. 
Depending on the location of the low energy synchrotron peak of the SED, blazars are sub-divided into low 
synchrotron peaked ($\nu_{peak}$ $<$ 10$^{14}$ Hz), intermediate synchrotron peaked (10$^{14}$ Hz $<$ $\nu_{peak}$ $<$ 10$^{15}$ Hz)
and high synchrotron peaked ($\nu_{peak}$ $>$ 10$^{15}$ Hz) 
blazars \citep{2010ApJ...716...30A}. While FSRQs are
predominantly low synchrotron peaked blazars, BL Lacs belong to all three classes with a majority of
them belonging to the high synchrotron peak category.

In the leptonic emission model of blazar jets, the high energy emission component
is interpreted as the inverse Compton scattering of low-energy photons
\citep{2010ApJ...716...30A}. The target low energy photons for inverse Compton emission 
can be the synchrotron photons, commonly referred as synchrotron self Compton or 
SSC \citep{1981ApJ...243..700K}, or the photons external to the jet called the external Compton or EC \citep{1987ApJ...322..650B}. 
Under this emission scenario,
one would expect to see a close correlation between optical and GeV flux 
variations since the same electron population is responsible for the emission at these energies. 
In the hadronic emission model,  the high energy component is interpreted as a result of hadronic
processes. Under this scenario 
one need not expect to see a correlation between optical and GeV flux 
variations since different species are responsible for the emission at these energies. However, systematic investigation of the correlation between
optical and GeV flux variations have led to varied results with instances
of (a) correlation between optical and GeV variations (b) instances
of optical flare without a GeV counterpart and (c) GeV flare without 
the corresponding optical counterpart \citep{2019MNRAS.486.1781R,2020MNRAS.498.5128R,2021MNRAS.504.1772R}. 
Thus, correlated studies of flux 
variations in the optical and GeV-band could not constrain the high energy 
emission process unambiguously. 

Similarly, the correlation observed between the radio and GeV $\gamma$-ray fluxes during various activity states support the region responsible for these emission to be co-spatial. The radio emission is due to synchrotron radiation by the relativistic electrons in the jet, while the GeV emission is due to the inverse Compton process. Interestingly, often radio variations are found to lag the GeV variations \citep{2014MNRAS.445..428M, 2014A&A...571L...2R, 2015A&A...576A.122E, 2022MNRAS.510..469K, 2023ApJ...953...47Y}. This lag may be associated with the slow cooling of the radio emitting electrons compared to the $\gamma$-ray. 

In the hadronic model of the high energy emission from blazar jets, 
the plausible emission mechanisms can be proton synchrotron \citep{2000NewA....5..377A} or 
hadronic cascades \citep{1993A&A...269...67M}. Accelerated protons
on interaction with cold protons or low energy photons in the surrounding 
medium could lead to the production of neutrinos as well as high 
energy $\gamma$-rays \citep{2023arXiv230106565P}. Additionally, there are now
increased evidence of blazars
being extragalactic neutrino sources \citep{2023arXiv230511263B,2023MNRAS.523.1799P}. 
The correlation found between the hard X-ray emission and neutrino emission
in blazars \citep{2023arXiv230600960P} tends to suggest that neutrinos are likely to come
from highly beamed blazars strong in X-rays. This led to arguments favoring
neutrinos being produced in proton-photon interaction.
The association of neutrinos with blazars has also revived the investigation 
on the hadronic origin of high energy emission from blazar jets.
 Therefore, a comparative
analysis of the SED characteristics of candidate neutrino blazars during the epoch of 
neutrino detection and quiescent/other flaring epochs not coincident with 
neutrino detection could provide the needed constraints on the
high energy emission process in blazars.

The first blazar found to be associated with the detection of neutrinos by the 
IceCube collaboration is TXS 0506+056 observed on 22 September 2017 \citep{2018Sci...361.1378I}. 
The detection of neutrinos  
was coincident in direction and time with the $\gamma$-ray flare from the 
source. Since then, few blazars have been found to be spatially coincident with the IceCube neutrino events \citep{2020ApJ...902...29P,2022arXiv220412242B,2023MNRAS.519.1396S}.
Generation of broadband SED of a large sample of neutrino blazars in a homogeneous 
manner and their systematic modeling can provide the key to enhancing our 
understanding of the high energy component in their broadband SED. We are 
carrying out such an investigation and in this work, we present our results
on the source PKS 0735+178.

The intermediate synchrotron peaked blazar,  PKS 0735+178 is one of the brightest BL Lac objects in the sky.
It was identified as a BL Lac first by \cite{1974ApJ...190L.101C} and is situated
at a  redshift of $z$ =  0.45 \citep{2001AJ....122..565R}. 
However, recently, \cite{2021ATel15132....1F} suggest a redshift of 
$z$ = 0.65.
It was detected in the $\gamma$-rays by the Energetic Gamma Ray
Experiment Telescope onboard the Compton Gamma Ray Observatory \citep{1999ApJS..123...79H}
and is also detected by the {\it Fermi} Gamma Ray Space 
Telescope, (hereafter Fermi; \cite{2020ApJS..247...33A}).  
Recently, it was found to be in spatial coincidence with 
multiple neutrino events by IceCube \citep{2021GCN.31191....1I}, 
Baikal \citep{2021ATel15112....1D}, Baksan \citep{2021ATel15143....1P}, and 
KM3NeT \citep{2022ATel15290....1F} detectors. This detection of 
neutrinos by multiple detectors was coincident with  
the largest flare ever observed in this source in the optical, UV, soft X-ray, and $\gamma$-ray bands. 
In this work, we carried out a detailed analysis of this source, including 
the variability, gamma-ray spectral study, and broad-band SED analysis with 
the help of multi-wavelength data. We also modeled the broadband  SEDs using simple one-zone 
leptonic model and the discrepancy of this model in explaining the $\gamma$-ray spectrum is inferred 
with the hadronic photo-meson process.  This paper is organized as follows: Multi-wavelength data and reduction is described
in Section \ref{sec:mwdr}, the analysis, results, and the broadband spectral fitting are discussed in Section \ref{sec:anl}
and the results of the work are summarized in Section \ref{sec:dis}.  Throughout this paper, we used the cosmological constants 
$\rm \Omega_M = 0.3$, $\rm \Omega_\Lambda = 0.7$, and $\rm H_0$ = 71  $km s^{-1} Mpc^{-1}$.

\section{Multi-wavelength data and reduction} 
\label{sec:mwdr}
The observational data used in this work was from the Large Area Telescope 
(LAT;\cite{2009ApJ...697.1071A}) on board the {\it Fermi}
Gamma-ray Space Telescope  over the period of about 14 years from August 2008 to February 2022.
In addition to the $\gamma$-ray data, we also used UV, optical, and X-ray data covering
the same period from the {\it Swift} telescope.

\subsection{$\gamma$-ray}
We used all data for PKS 0735+178 collected for the period 2008 August to 2022 February (MJD: 54749$-$59611) from the Fermi archives. We utilised the {\it{fermipy}} package{\footnote{https://fermipy.readthedocs.io/en/latest/}}, a python package that facilitates analysis of data from the LAT with the Fermi Science Tools for the generation of spectral points. We extracted photon-like events categorized as 'evclass=128, evtype=3' with 
energies 0.1$\leqslant$E$\leqslant$300 GeV $\gamma$-rays within a circular region of interest (ROI) of 15$^\circ$ centered on the source. We applied the recent isotropic model, "iso$\_$P8R2$\_$SOURCE$\_$V6$\_$v06" 
and the Galactic diffuse emission model "gll$\_$iem$\_$v06". For generation of the $\gamma$-ray light curves, we considered
the source to be detected if the test statistics (TS) > 9, which corresponds to a 3$\sigma$ detection \citep{1996ApJ...461..396M}. For bins with TS $<$ 9, we considered the source as undetected.

\subsection{X-ray}
We used data from the {\it Swift} X-ray Telescope (XRT) collected from the 
HEASARC{\footnote{https://heasarc.gsfc.nasa.gov/docs/archive.html}} archives for 
X-rays with energies between 0.3 $-$ 10 keV from August 2008 to February 2022. We reduced the data using the default 
parameter values in accordance with the instructions provided by the instrument 
team. The source spectra were 
chosen from a region of radii 60", whereas the background spectra were taken from 
a region of radii 120". We combined the exposure map using the tool XIMAGE and 
created the ancillary response files with xrtmkarf. We used an absorbed simple 
power law model with the Galactic neutral hydrogen column density of 
3.74 $\times$10$^{20}$ cm$^{-2}$ to perform the fitting within XSPEC \citep{1996ASPC..101...17A}.

\subsection{UV and Optical}

For the analysis of UV and optical data from August 2008 to February 2022, we utilized the data 
obtained from the {\it Swift}-UV-Optical telescope (UVOT), an instrument onboard 
the {\it Swift} spacecraft. We used the data from the filters V, U, W1 and W2. The central wavelength(FWHM) of these filters are 5468 \AA(796 \AA), 3465 \AA (785 \AA), 2600 \AA (693 \AA) and
1928 \AA (657 \AA) \citep{2008MNRAS.383..627P}.
We processed the data from {\it Swift}-UVOT using the 
online tool provided by the telescope's data archive. In order to account for 
the effects of galactic absorption, we applied corrections to the UV and optical 
data points during SED analysis.

\section{Analysis and Results}
\label{sec:anl}
\subsection{Multi-wavelength light curve}
\label{sec:ml} 
\begin{figure*}
\hspace*{-2.0cm}\includegraphics[scale=0.44]{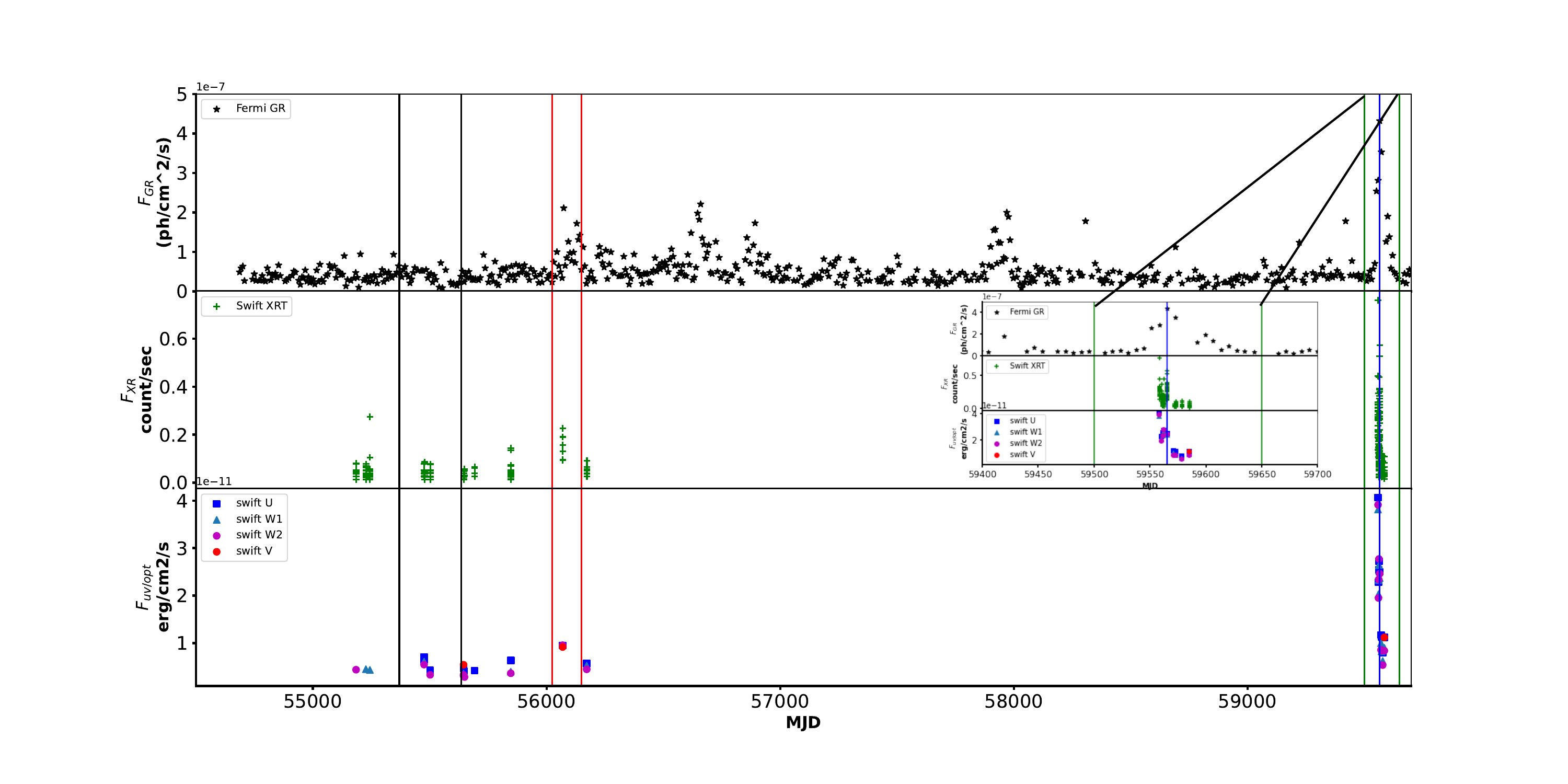}
\caption{Multi wavelength light curves of the source PKS0735+178. From the top, the 
first panel shows the weekly binned $\gamma$-ray light curve for the time range 
MJD 54749$-$59611; the second panel shows the SWIFT-XRT light curve in photon counting mode, and the third panel shows the  Swift UVOT light curves in W1, W2, U, and V-bands. The blue vertical line is the epoch of neutrino detection. The region encompassed by the black vertical lines,
red vertical lines and green vertical lines refer to the epochs $E_{1}$, $E_{2}$ and $E_{3}$. The inset is the zoomed region around the flaring epoch $E_{3}$.}
\label{figure-1}
\end{figure*}

We utilized the one week binned $\gamma$-ray light curve in the 100 MeV to 300 GeV band generated using the
procedure outlined in Section 2.1. 
The  multi-wavelength light curves spanning about 14 years that include 
$\gamma$-ray, X-ray, UV, and optical from 2008 August to 2022 
February (MJD: 54749$-$59611) are shown in Fig \ref{figure-1}. From Fig \ref{figure-1}, 
it is evident that PKS 0735+178 has gone through both quiescent and active phases. 
During this period, we identified three time intervals. They are denoted as 
epochs $E_{1}$, $E_{2}$, and $E_{3}$.  $E_{1}$ covers the period MJD 55370$-$55635 (265 days), 
when the source was in a quiescent state, $E_{2}$ covers the period  MJD 56025$-$56150 (125 days),
when the source was in an active state and $E_{3}$ covers the period MJD 59500$-$59650 (150 days) 
when the source was in a historically high state coincident with the epoch of 
neutrino detection. The neutrino event detected on December 8, 2021 (MJD 59556), 
is marked as a blue vertical line in Fig \ref{figure-1} in $E_{3}$. During this
epoch, the source showed the largest flare ever observed in the optical, UV, 
X-rays and $\gamma$-rays. 

\subsection{Variability Analysis}
\subsubsection{Long term variability}
To quantify the variability of the source on time scales of weeks, we calculated the 
fractional variability amplitude (F$_{var}$) in different energy bands following 
\cite{2003MNRAS.345.1271V}. We define 

\begin{equation}
    \label{eq3}
    F_{var}=\sqrt{\frac{S^2-\overline{\sigma_{err}^2}}{\overline{x}^2}}
\end{equation}
where $\rm S^2$ is the variance, $\rm \overline{x}$ is the mean and $\rm \overline{\sigma_{err}^2}$ is the mean square of the measurement error on the flux points. The uncertainty on $\rm F_{var}$ is  
given by \citep{2003MNRAS.345.1271V}

\begin{equation}
    \label{eq4}
    F_{var,err}=\sqrt{\frac{1}{2N}\left(\frac{\overline{\sigma_{err}^2}}{F_{var}\overline{x}^2}\right)^2+\frac{1}{N}\frac{\overline{\sigma_{err}^2}}{\overline{x}^2}}
\end{equation}

Here, N represents the number of flux points in the light curve. We calculated 
the $\gamma$-ray variability for $E_{1}$, $E_{2}$, $E_{3}$ and for the total 
light curve. The results of the variability analysis are given in  Table~\ref{table-1}.

\begin{table}
\centering
\caption{{The $\gamma$-ray fractional variability amplitude ($\rm F_{var}$) in different epochs in the 100 MeV to 300 GeV band. Here, $E_{1}$ corresponds to the quiescent state, $E_{2}$ corresponds to the active state  and $E_{3}$ corresponds to the high state of the source that coincides with the neutrino detection.}}
\begin{tabular}{l r}
\hline 
Epoch  & $\rm F_{var}(\%)$ \\
\hline
$E_{1}$  & --- \\
$E_{2}$  & 41.3$\pm$ 7.9\\
$E_{3}$ & 95.2$\pm$ 6.5\\
Total LC & 61.9$\pm$ 4.3\\

\hline
\end{tabular}
\label{table-1}
\end{table}

\subsubsection{Short term variability}
Blazars are known to show $\gamma$-ray flux variations on time scales lesser than an hour.
\citep{2011A&A...530A..77F,2013ApJ...766L..11S,2022A&A...668A.152P}.
Detection of such short time scale variation could enable one to constrain the size and
the location of the $\gamma$-ray emission region. We, therefore, searched for the 
presence of flux variations on very short time scales (of the order of hours) in the 
$\gamma$-ray light curve of PKS 0735+178. For this, we identified the epoch of 
very high $\gamma$-ray activity, namely, $E_{3}$  and generated one day binned light curves. We then calculated the flux doubling/halving time scale
using the relation
\begin{equation}
F(t) = F(t_0) \times 2^{-(t - t_0)/\tau}
\end{equation}
Here F(t) and  F(t$_0$) are the fluxes at time t and $t_0$ 
respectively and $\tau$ is the flux doubling/halving time scale.
This calculation was done with the condition that the 
flux difference between epochs t and t$_0$ is greater than
3$\sigma$ \citep{2011A&A...530A..77F}.
Using the flux doubling time scale, we constrained the size of the $\gamma$-ray emitting region as
\begin{equation}
	R \leq \frac{c\tau\delta}{1+z}
\end{equation}
where $\delta$ is the Doppler Factor and $\tau$ is the flux doubling/having time-scale and c is the speed of light.

\subsubsection{Spectral Variability}
To investigate spectral variations in the source, during the epochs analyzed in 
this work, we followed a model-independent approach of estimating the hardness 
ratio (HR) and then investigated the dependence of HR on the total flux of
the source.  For this, we generated one day binned $\gamma$-ray light curves in two 
energy ranges, namely, 0.1$-$10 GeV and 10$-$300 GeV.  We define HR and the 
associated errors in HR  as

\begin{equation}
\label{eq6}
HR= \left(\frac{H-S}{H+S}\right)
\end{equation}

\begin{equation}
\label{eq7}
\sigma_{HR}= \frac{2}{(H+S)^2} \sqrt{H^2\sigma_S^2+S^2\sigma_H^2}
\end{equation}

\begin{figure*}
\begin{center}
\vbox
   {
    \hbox{
         \includegraphics[width=90mm,height=50mm]{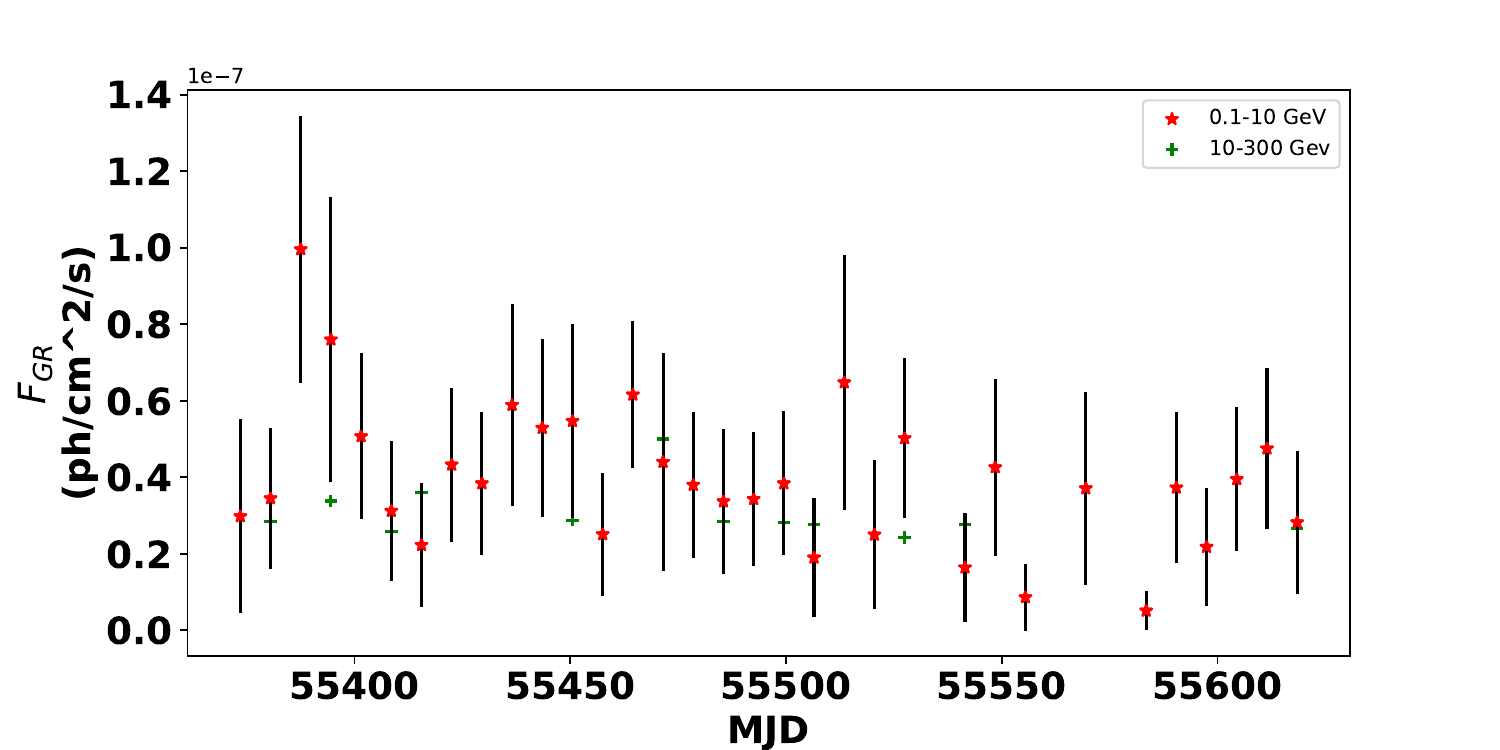}
        \includegraphics[width=90mm,height=50mm]{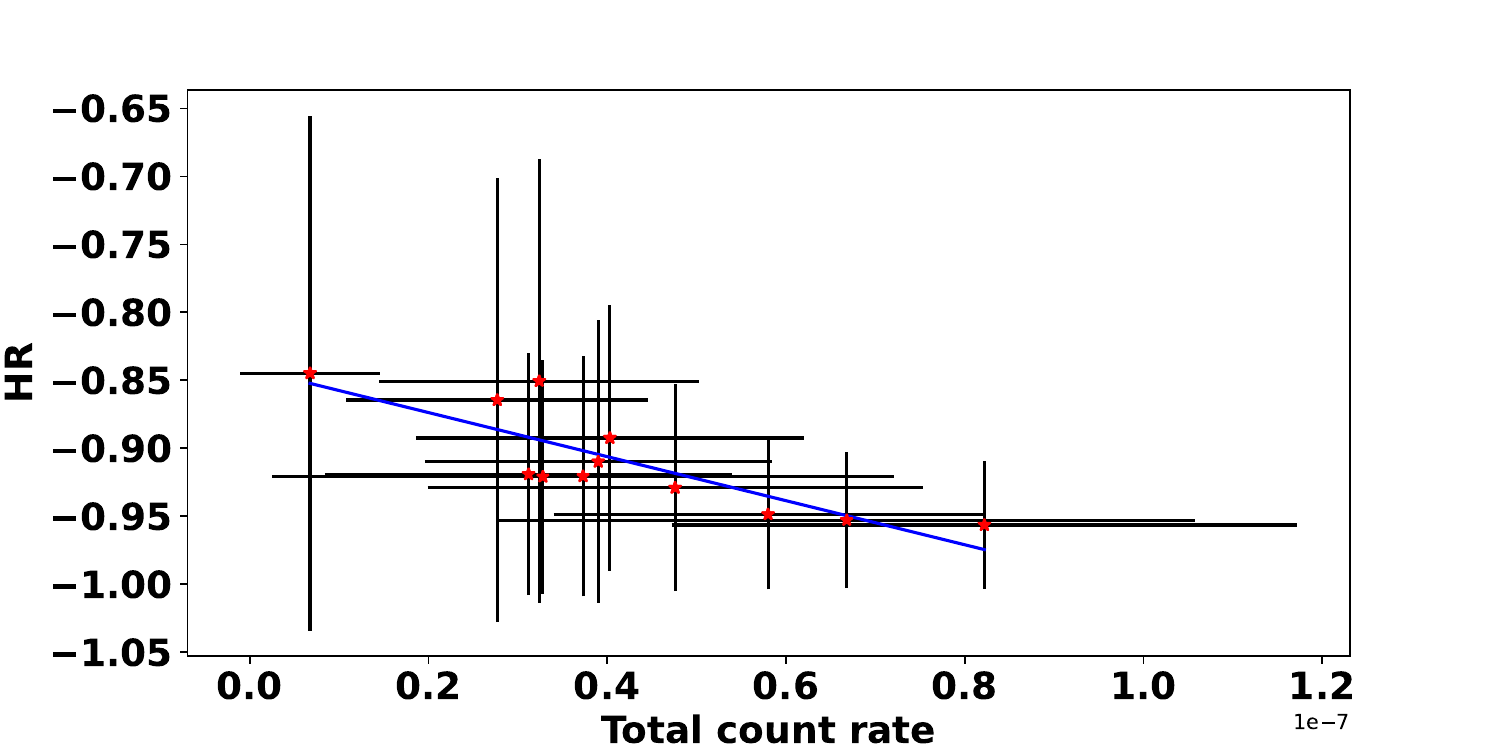}

         }
    \hbox{
         \includegraphics[width=90mm,height=50mm]{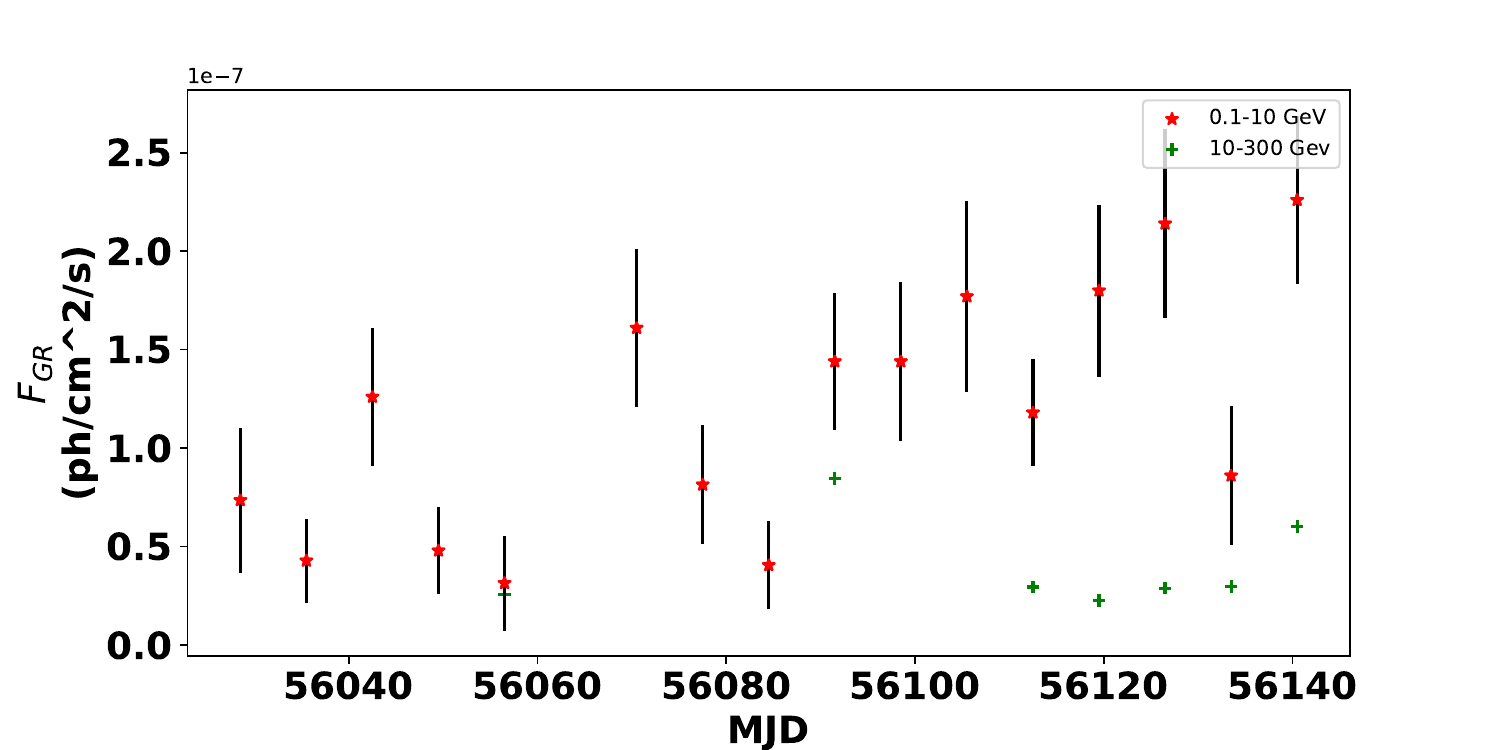}
         \includegraphics[width=90mm,height=50mm]{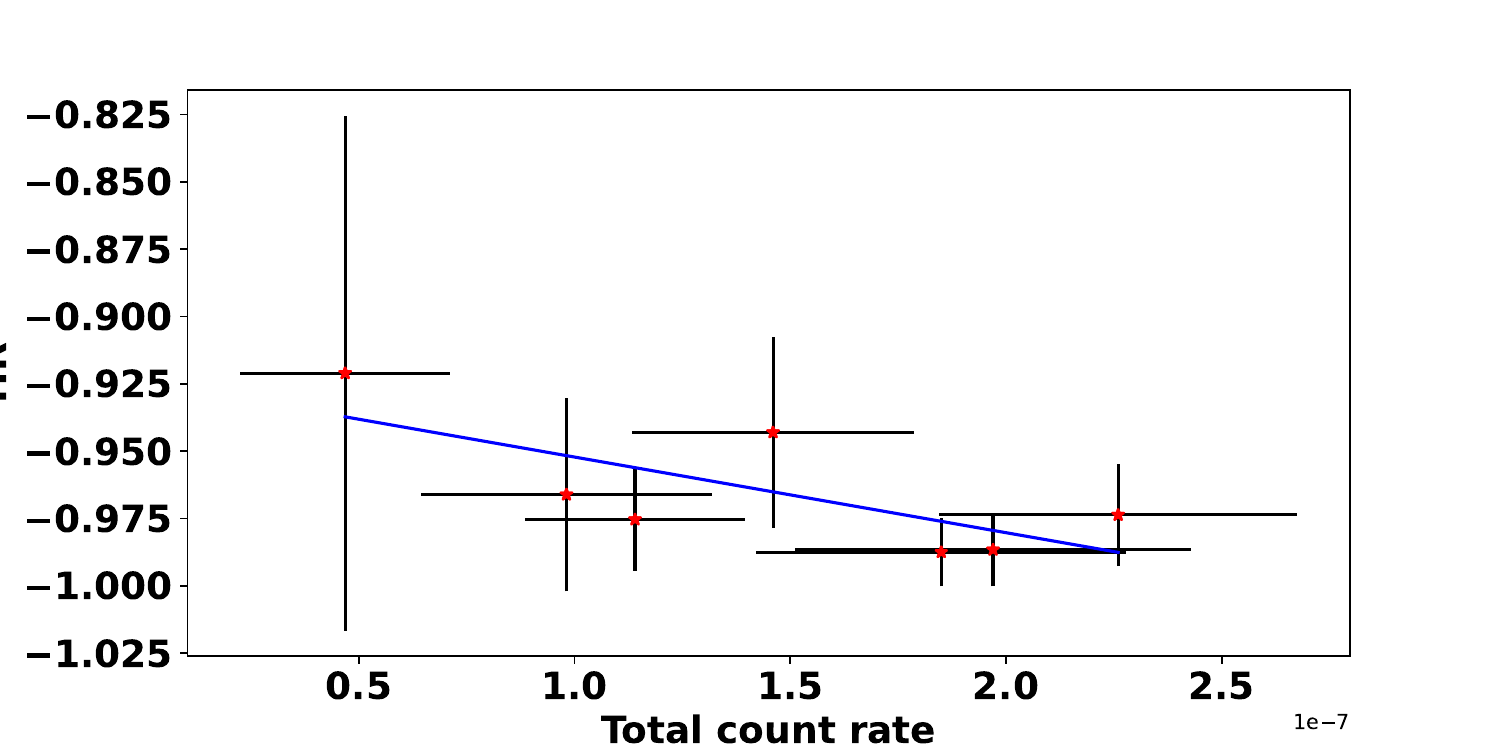}
         }
    \hbox{
         \includegraphics[width=90mm,height=50mm]{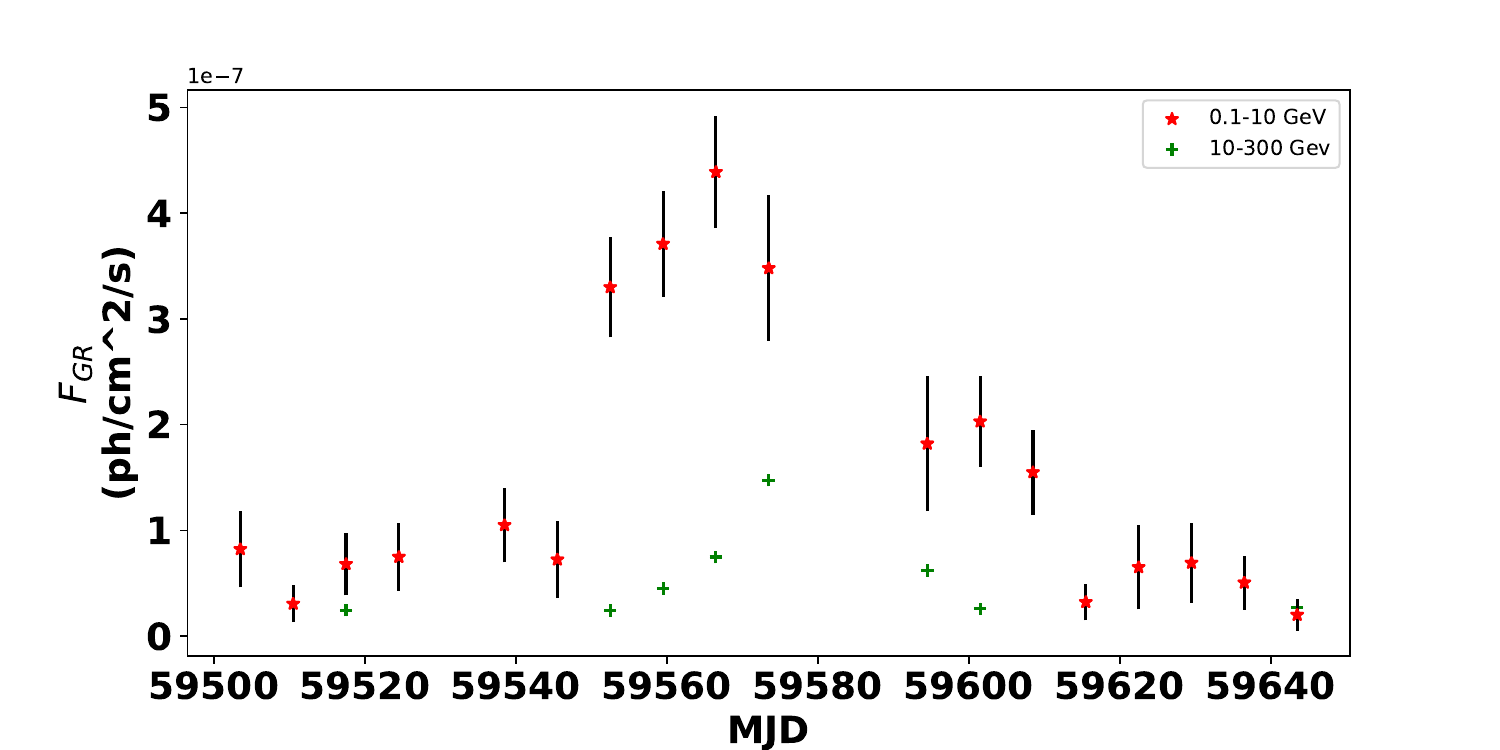}
         \includegraphics[width=90mm,height=50mm]{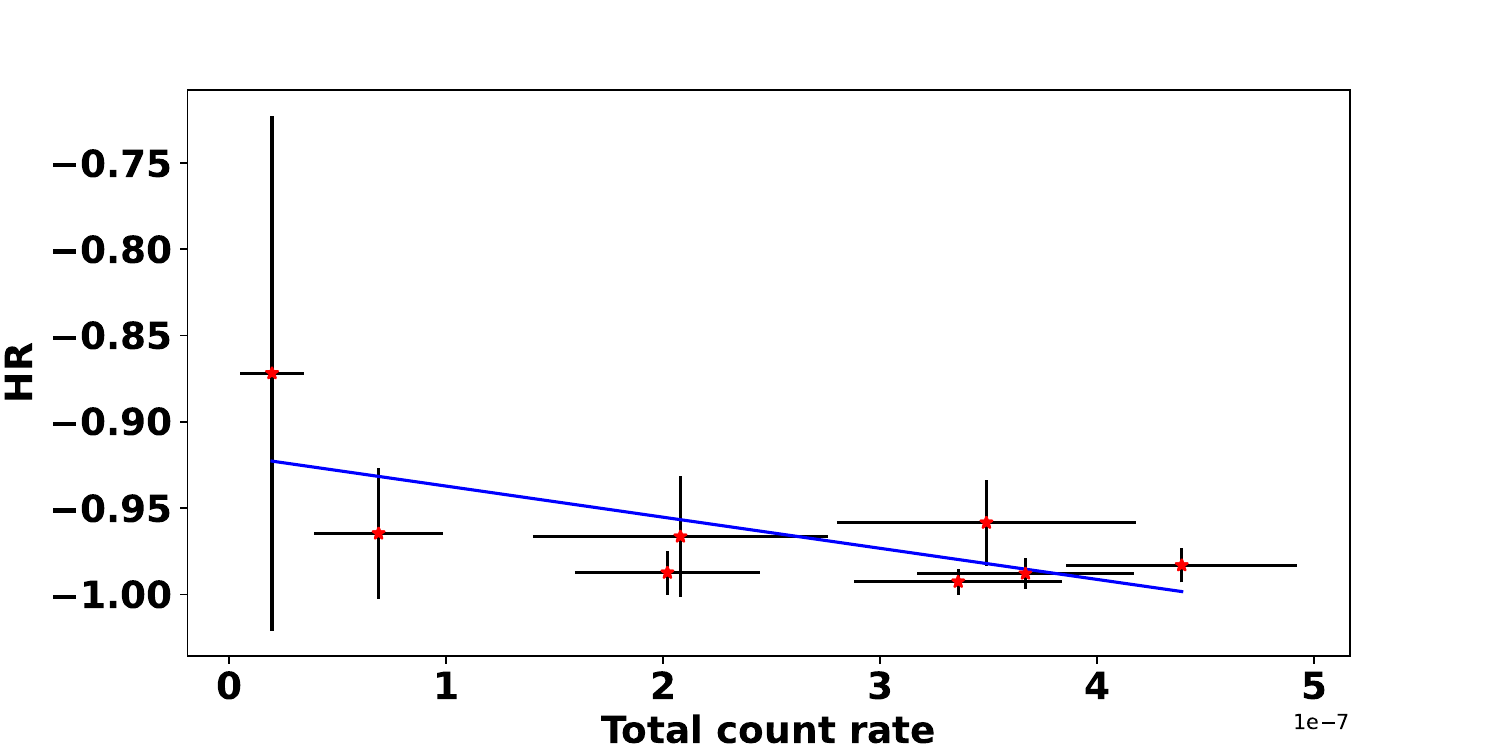}
        }
   } 
\caption{Left panels: One week binned gamma-ray light curves for the
epochs $E_{1}$ (top left), $E_{2}$ (middle left) and $E_{3}$ (bottom left) in the 0.1 - 10 GeV (red) and 10$-$300 GeV multiplied by a factor of 20 (green). Right panels: Hardness ratio versus total intensity in the 0.1 - 300 Gev band for the Epochs $E_{1}$ (top right), $E_{2}$ (middle right) and $E_{3}$ (bottom right).  Here the solid blue lines are
the weighted linear least squares fit to the data.}
\label{figure-2}
\end{center}
\end{figure*}

The generated light curves in the two energy ranges and the corresponding 
HR for the three epochs are shown in Fig. \ref{figure-2}. From the light curves, it is 
evident that the hard flux counts are lower compared to the soft flux counts. 
We carried out a weighted linear least square fit to the HR v/s intensity diagram
in Fig. \ref{figure-2}. The results of the fit are given in Table \ref{table-2}. 
We found significant evidence of spectral variation in only one epoch, namely $E_{1}$, wherein we found a softer when brighter trend. This is at odds with the recent finding of a harder when brighter
trend seen in the $\gamma$-ray band \citep{2022ApJ...933..224F}.
Close to the period of neutrino detection, in the X-ray band
the source was found to show a softer when brighter trend, generally
seen in the FSRQ category of blazars \citep{2023arXiv230106565P}.

\begin{table}
\centering
\caption{Results of the weighted linear least squares fit to the HR v/s total intensity diagram. Here, the epochs $E_{1}$, $E_{2}$ and $E_{3}$ correspond to the quiescent, active and high state of the source, R is the linear correlation coefficient and p is the probability for no correlation.}
\label{table-2}
\begin{tabular}{lccrc} 
\hline
Epochs  & slope ($\times$$10^{7}$) & intercept  & R & p \\ \hline
$E_{1}$      &   $-$0.16$\pm$0.01     &     $-$0.84$\pm$0.02        &   $-$0.83 &  0.001 \\  
$E_{2}$      &    $-$0.03$\pm$0.01    &      $-$0.92$\pm$0.02       &   $-$0.72 &  0.066 \\
$E_{3}$      &    $-$0.02$\pm$0.00   &      $-$0.92$\pm$0.02    &   $-$0.69  &  0.061 \\ \hline
\end{tabular}
\end{table}

\subsection{$\gamma$-ray spectra}
\begin{figure}
\begin{center}
\vbox{
     \includegraphics[scale=0.42]{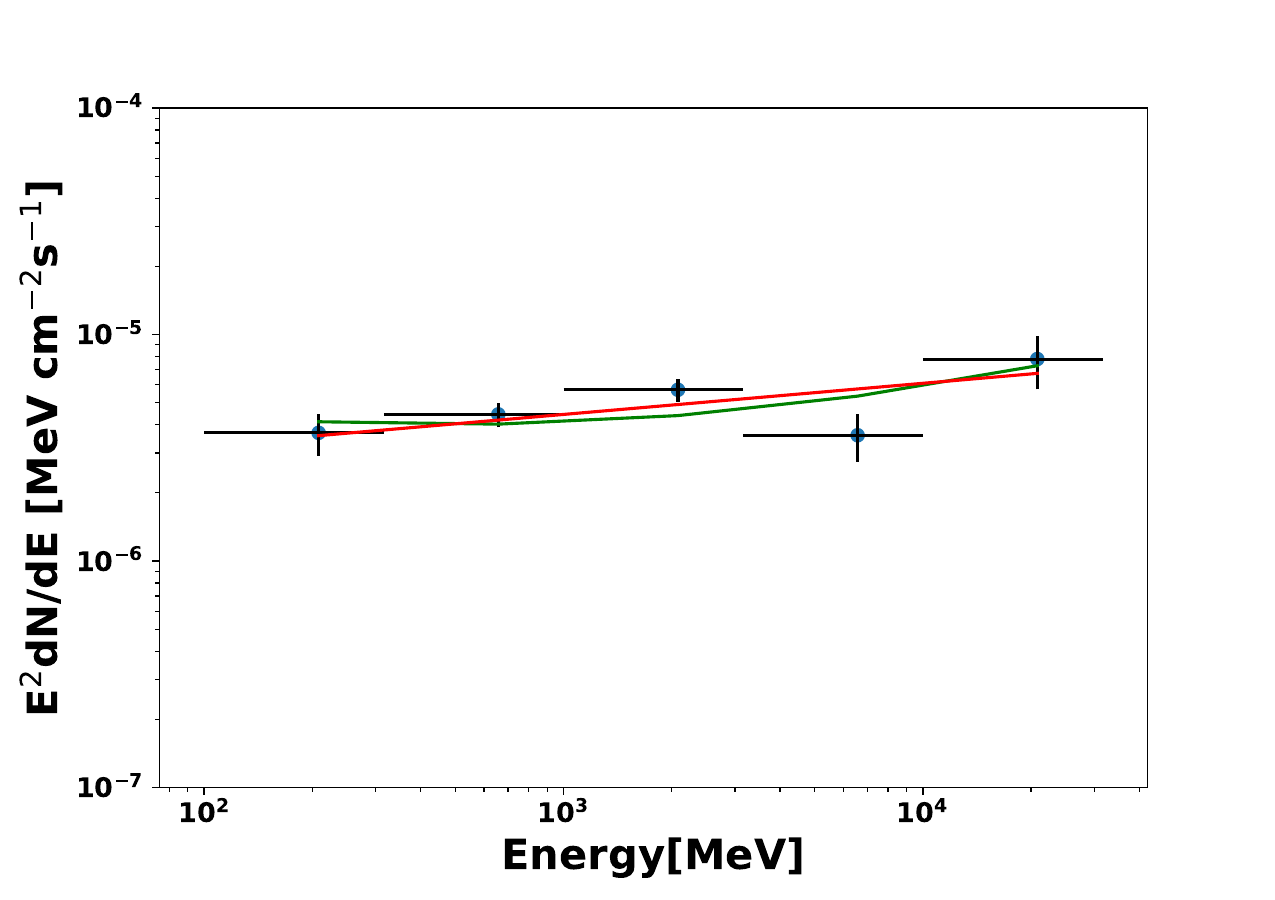}
     \includegraphics[scale=0.42]{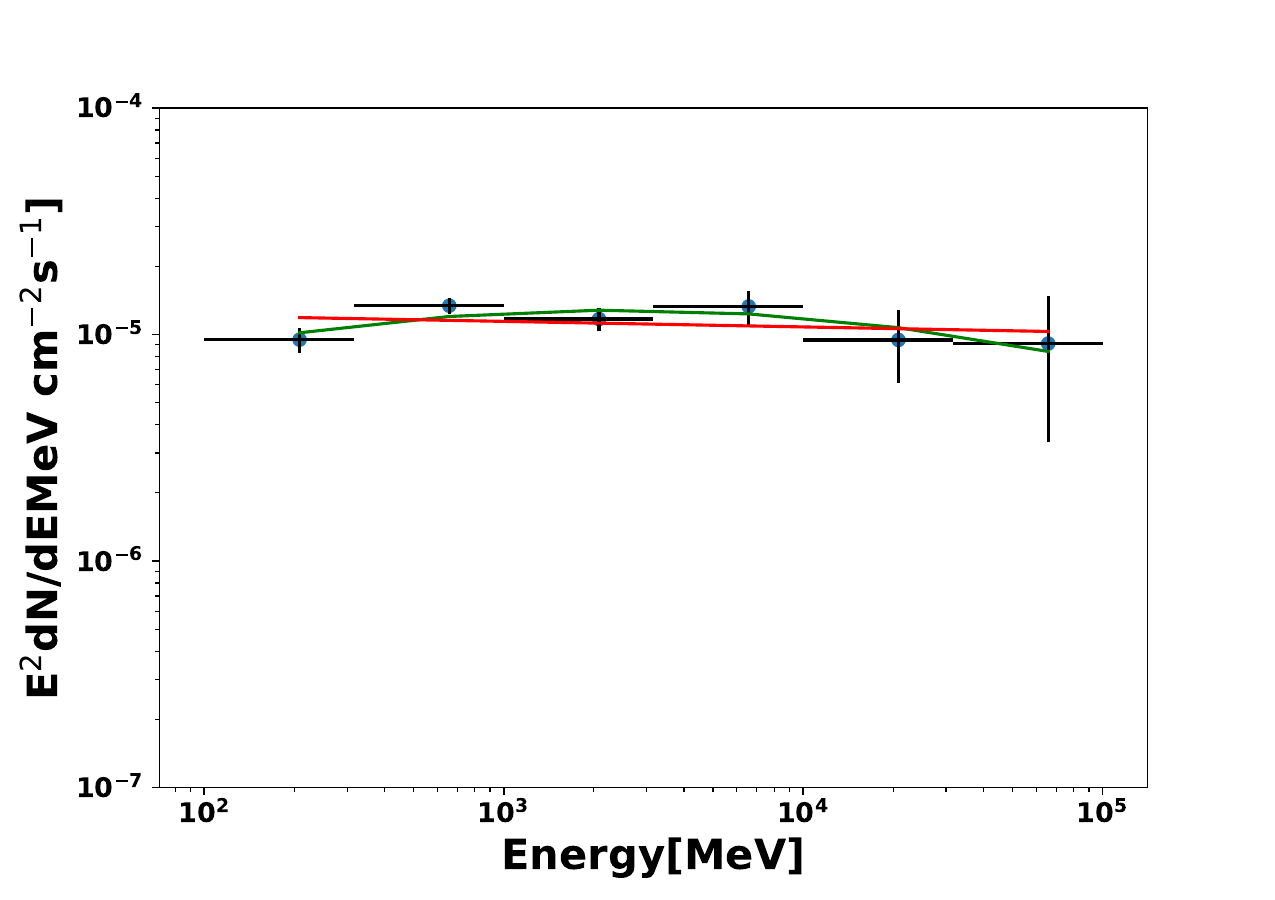}
     \includegraphics[scale=0.42]{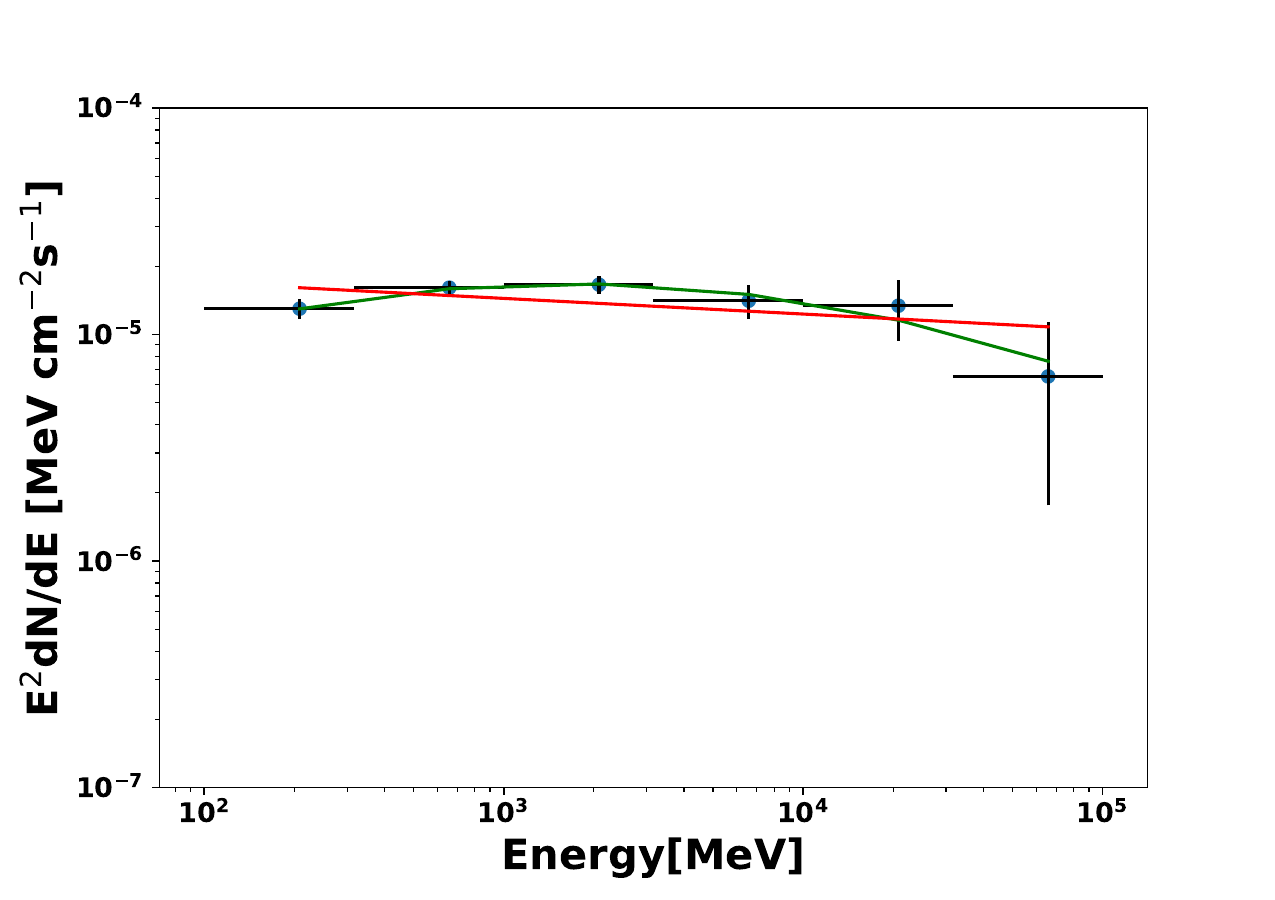}
      }
\end{center}
\caption{Simple power law (red solid line) and log parabola (green solid line)  fits to the 
$\gamma$-ray spectra of PKS 0735+178 during epochs $E_{1}$ (top panel), $E_{2}$ (middle panel) and $E_{3}$ (bottom panel).}
\label{figure-3}
\end{figure}

The slope of the $\gamma$-ray spectrum can provide hints on the intrinsic 
accelerated electron distribution responsible for this emission process.
We carried out 
the analysis of the $\gamma$-ray spectra of the three epochs using two
spectral models namely the power law (PL) and log parabola (LP) models. 
The PL model has the functional form as

\begin{equation}
\label{eq8}
\frac{dN(E)}{dE} = N_0 E^{-\Gamma}
\end{equation}
Here, dN(E)/dE is the differential photon number (cm$^{-2}$ s$^{-1}$ MeV$^{-1}$), 
$\rm N_0$ is the normalisation and $\Gamma$ is the photon index. 
The LP model is defined as \citep{2012ApJS..199...31N}
\begin{equation}
dN(E)/dE=N_{\circ}(E/E_{\circ})^{-\alpha-\beta ln(E/E_{\circ})}
\end{equation}
where, $\alpha$ is photon spectral slope at energy $E_{\circ}$ and $\beta$ defines the peak
spectral curvature of the SED.
We used the Maximum Likelihood estimator {\it gtlike} and the 
likelihood ratio test \citep{1996ApJ...461..396M} to check the PL model (null hypothesis) 
against the LP model (alternative hypothesis). We calculated the 
curvature of test statistics $TS_{curve}$ = 2(log $L_{LP}$ - log $L_{PL}$) 
following \cite{2012ApJS..199...31N}. We tested the 
presence of a significant curvature by setting the condition 
$TS_{curve}$ $>$ 16.
The $\gamma$-ray spectra along with the model fits for the epochs $E_{1}$, $E_{2}$ and $E_{3}$ 
are shown  in Fig \ref{figure-3}. The parameters of the fit are given in Table 
~\ref{table-3}. Significant curvature in the $\gamma$-ray spectrum is not
found in the three epochs.

\begin{table*}
\centering
\caption{Results of the power law (PL) and log parabola (LP) model fits
to the three epochs $E_{1}$, $E_{2}$ and $E_{3}$. Here, $E_{1}$, $E_{2}$ and $E_{3}$ represent the quiescent,
active and high state of the source. Here, Gamma is the photon index, Flux is the
$\gamma$-ray flux value in units of $10^{-12}$ph $cm^{-2}$ $s^{-1}$, TS is the test
statistics, $\alpha$ is the spectral index, $\beta$ is the curvature in the spectra and $TS_{curve}$
signifies the presence of curvature in the spectra.}
\label{table-3}
\begin{tabular}{lccrccccccc} \hline
Epochs  & \multicolumn{4}{c}{PL} & \multicolumn{6}{c}{LP} \\ \hline
        & $\Gamma$  & Flux  & TS  & L$_{PL}$   & $\alpha$  & $\beta$  & Flux  & TS  & L$_{LP}$  & TS$_{curve}$ \\
\hline
$E_{1}$ &  $-$1.99 & 2.006 & 637.221 & -44224.847 & 1.94 $\pm$ 0.07 & 0.07 $\pm$ 0.04 &  2.162  & 634.591  &  -44225.597 &  $-$1.5 \\
$E_{2}$ &   $-$2.01 & 5.234 & 1050.448 & -26527.641 & 1.99 $\pm$ 0.05 & 0.07 $\pm$ 0.03 & 5.887   & 958.9225  & -26524.141  & 7.00 \\
$E_{3}$ &   $-$2.01 & 6.453 & 1381.112 & -25249.253 & 2.01 $\pm$ 0.05 & 0.07 $\pm$ 0.03 &  7.319  &  1359.155 & -25246.537  & 5.43 \\ \hline
\end{tabular}
\end{table*}

\subsection{Broad band SED analysis}
\begin{figure*}
\begin{center}
\vbox{
     \hbox{
             \includegraphics[width=80mm,height=60mm]{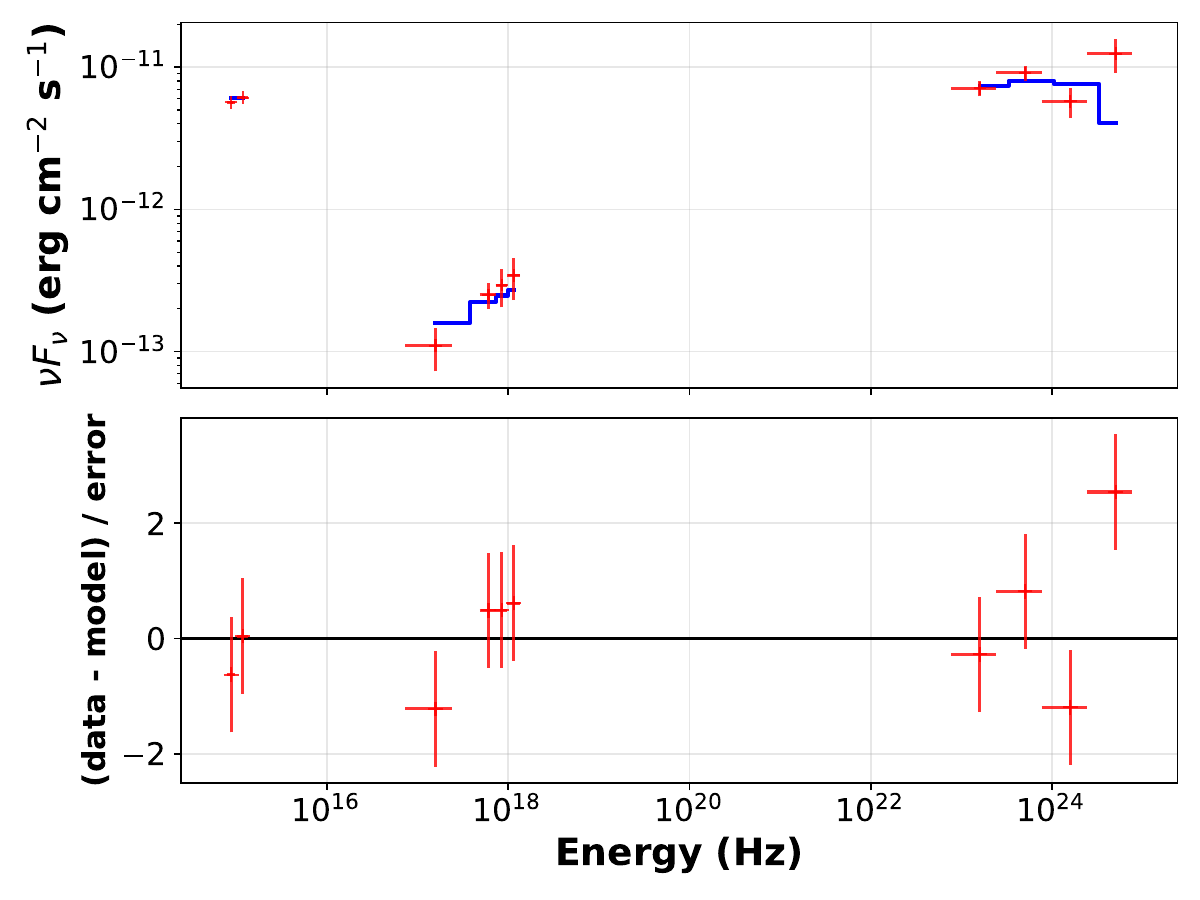}
              \includegraphics[width=90mm,height=60mm]{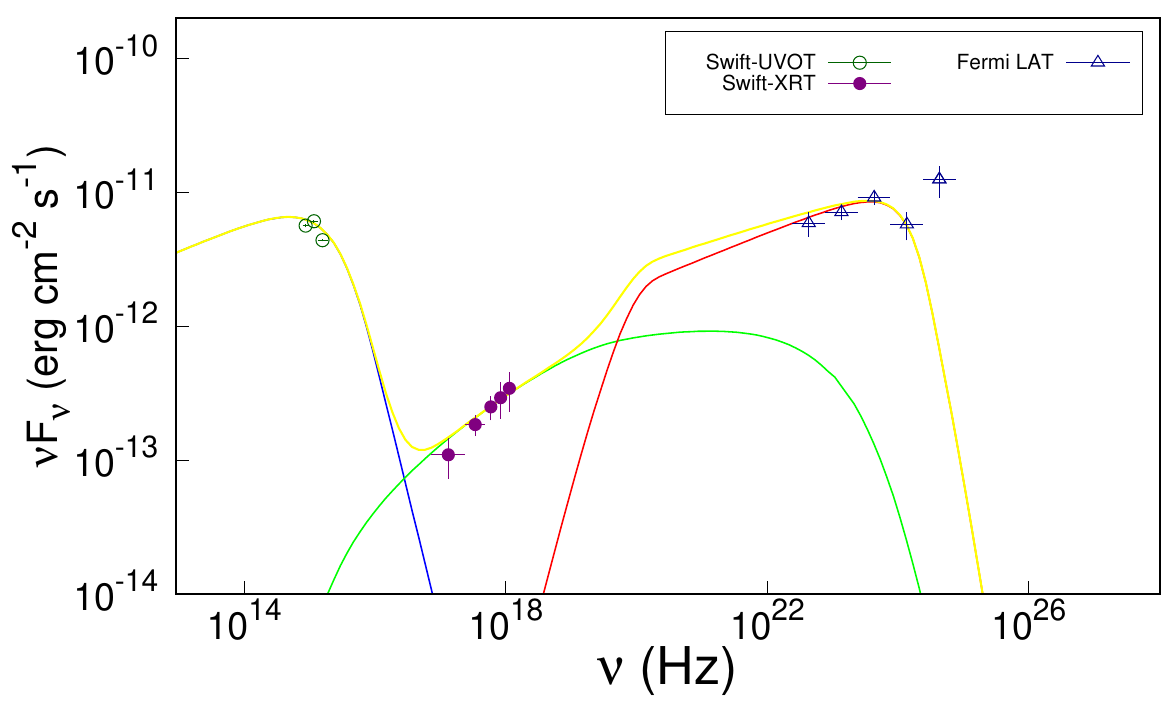}
          }
     \hbox{
            \includegraphics[width=80mm,height=60mm]{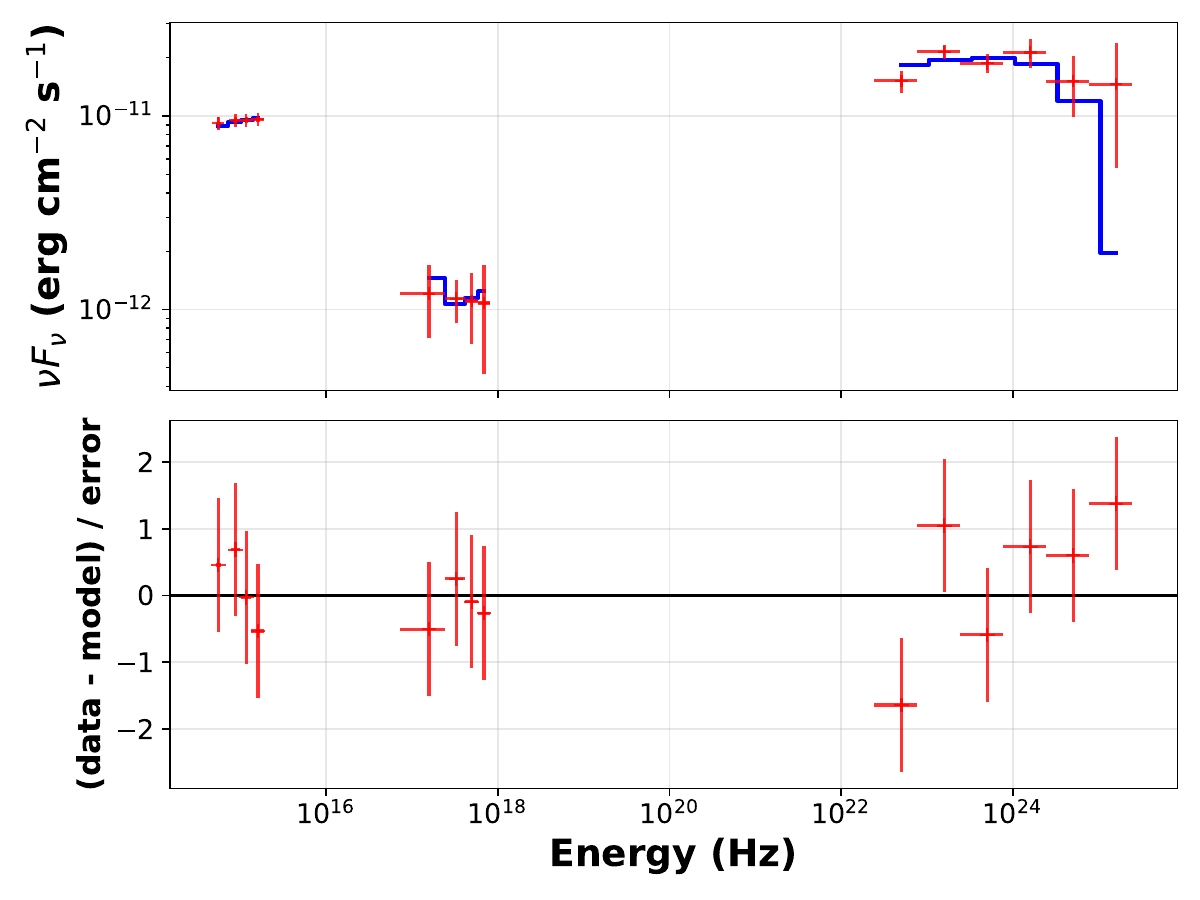}
             \includegraphics[width=90mm,height=60mm]{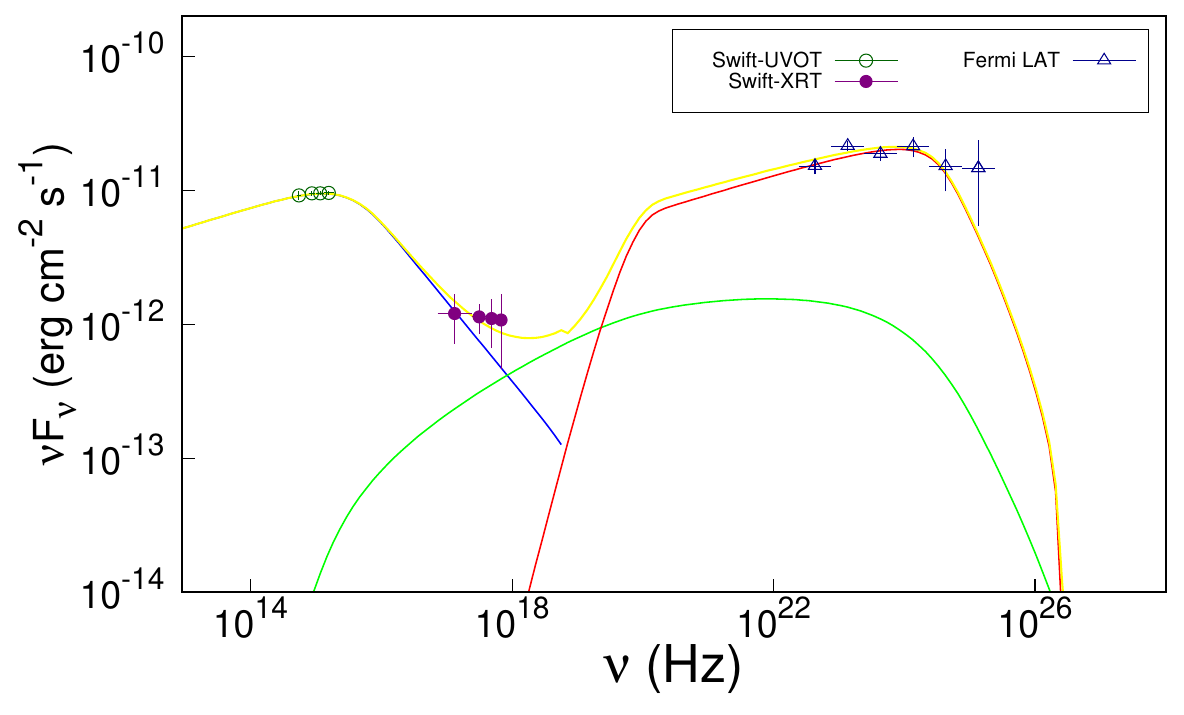}
          }
     \hbox{
             \includegraphics[width=80mm,height=60mm]{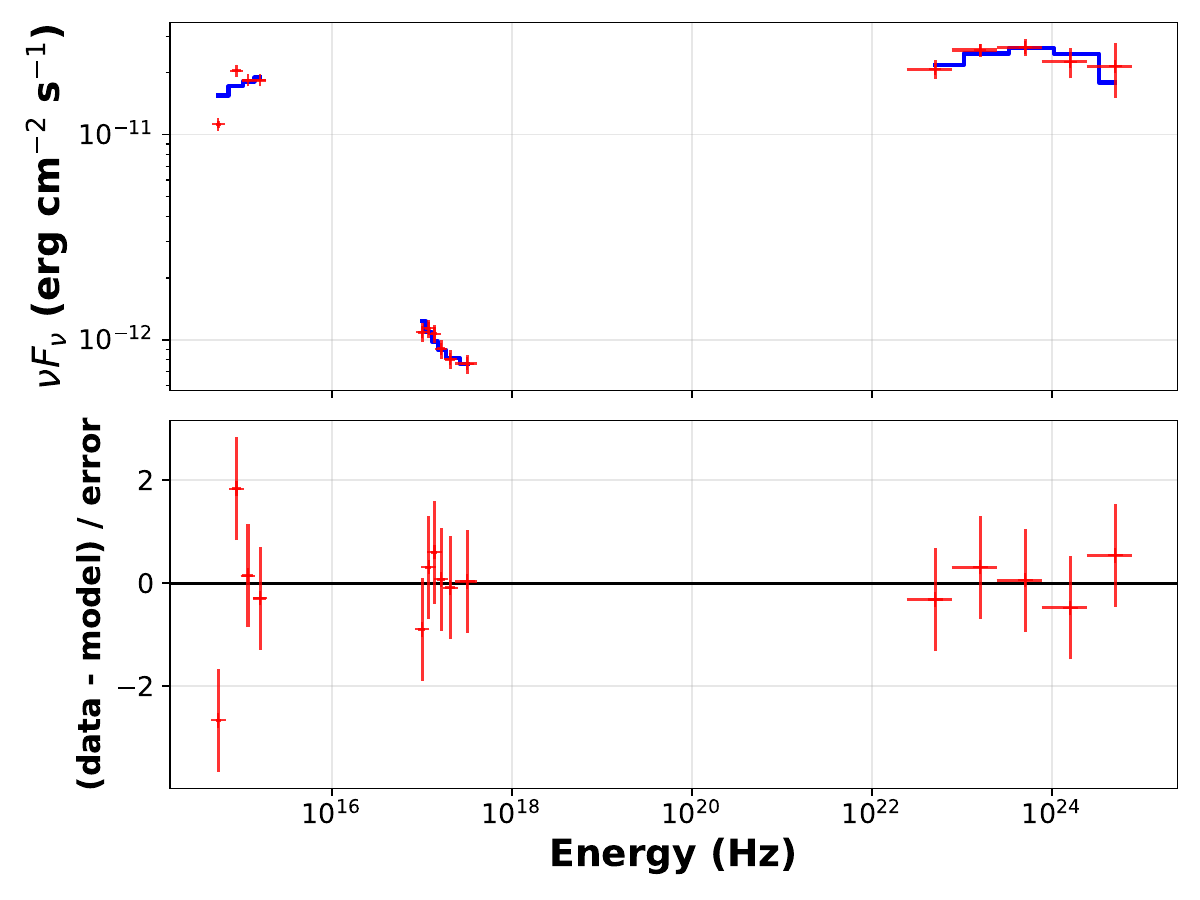}
            \includegraphics[width=90mm,height=60mm]{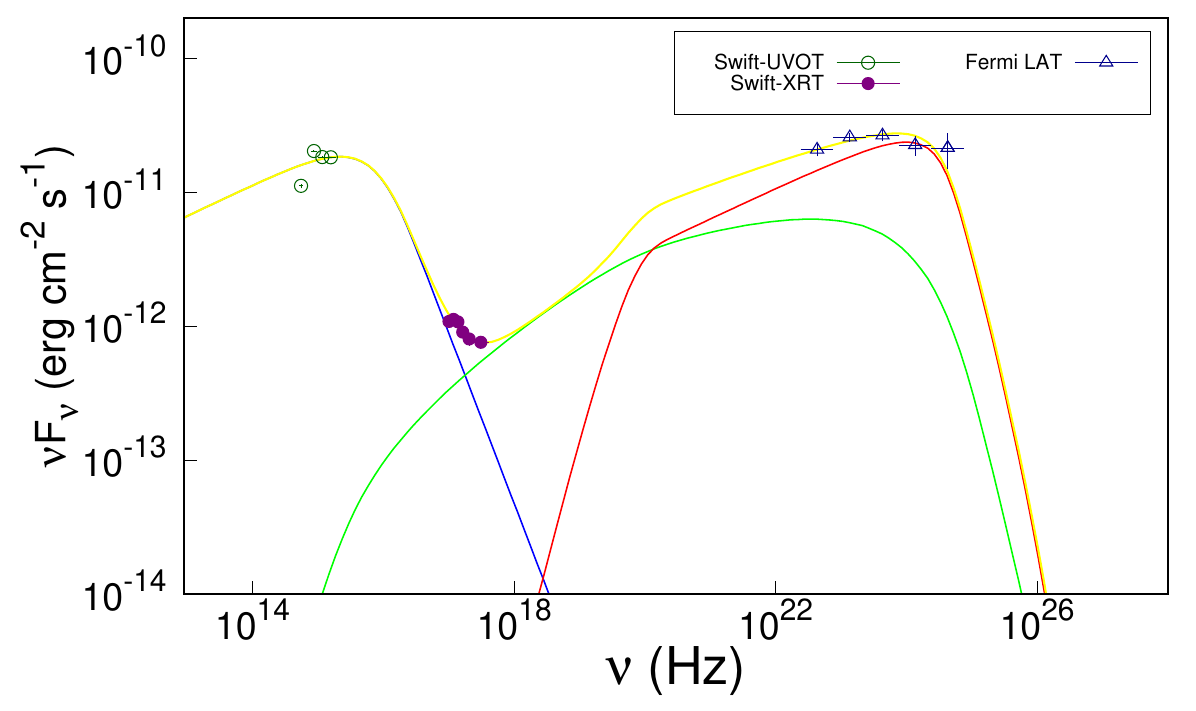}
          }
     }
\caption{Broad band spectral energy distribution along with the one zone leptonic emission model fits for $E_{1}$ (top panel), $E_{2}$ (middle panel) and $E_{3}$ (bottom panel). In the right panels the blue line refers to the synchrotron model, the green line refers to the SSC process and the red line refers to the EC process. The yellow line is the sum of all the components. In the left-hand panels for each epoch, the first panel shows the fitting of the model to the data and the second panel shows the residuals carried out in XSPEC.}
\label{figure-4}
\end{center}
\end{figure*}

We modeled the generated SEDs of the three epochs using the one zone leptonic 
model \citep{2018RAA....18...35S}. In this model, the emission region was 
assumed to be a spherical blob of size $R$ filled with non-thermal electrons 
following a broken power law distribution
\begin{align} \label{eq:broken}
	N(\gamma)\,d\gamma = \left\{
\begin{array}{ll}
	K\,\gamma^{-p}\,d\gamma&\textrm{for}\quad \mbox {~$\gamma_{\rm min}<\gamma<\gamma_b$~} \\
	K\,\gamma_b^{q-p}\gamma^{-q}\,d\gamma&\textrm{for}\quad \mbox {~$\gamma_b<\gamma<\gamma_{\rm max}$~}
\end{array}
\right.
\end{align} 
where, $\gamma$ is the electron Lorentz factor and, $p$ and $q$ are the low and 
high energy power-law indices with $\gamma_b$ the Lorentz factor corresponding 
to the break energy. The emission region is permeated with a tangled magnetic 
field $B$ and moves down the jet with a bulk Lorentz factor $\Gamma$. The broadband 
SEDs were modeled using synchrotron, SSC, and EC emission mechanisms. XSPEC
modeled SEDs of the source PKS 0735+178 for the selected epochs $E_{1}$, $E_{2}$ and $E_{3}$ are 
shown in Fig. \ref{figure-4}.  The numerical model is coupled as a local 
model in X-ray spectral fitting package XSPEC \citep{1996ASPC..101...17A} and the fitting was 
performed. We found from the fitting, that the $\gamma$-ray emission is better 
explained by the EC process with the target photons from the IR dusty torus
at a temperature of 1000 K. The best fit parameters obtained are 
listed in Table~\ref{table-4}. The best fit model along with the observed
data are shown in Fig \ref{figure-4}. Though classified as a BL Lac object, the broadband SED of PKS 0735+178 resembles that of a FSRQ with the requirement of EC
process to explain the observed high energy emission. This is also similar to the case of the first high energy neutrino blazar
TXS 0506+056, which was found not a BL Lac but a FSRQ \citep{10.1093/mnrasl/slz011}. The requirement of the external
target photon field to explain the observed SED in PKS 0735+178 as found here has also been hinted recently \citep{2023MNRAS.519.1396S,2023ApJ...954...70A}.

For all the epochs except $E_{1}$, the spectral
fit satisfies all the $\gamma$-ray flux points. The highest $\gamma$-ray flux point of
$E_{1}$ corresponding to energy 21 GeV cannot be explained under the leptonic model and 
we investigated this by considering the additional emission from photo-meson
process. In this lepto-hadronic model, we considered the interaction of
relativistic protons with the synchrotron photons.

We considered a particle distribution involving a combination of leptons and hadrons to model the excess high energy emission in $E_{1}$. The contribution of these protons is sub-dominant in the lower energy region. We considered a photo-meson process in which relativistic protons interact with the synchrotron photons. In this model, we only considered the production of $\pi^0$   meson and its decay into two $\gamma$ photons. We used the model developed by \cite{2008PhRvD..78c4013K}to incorporate the photo-meson emission process in the modeling of the SED of $E_{1}$. The energy distribution of $\gamma$-rays produced by the decay of $\pi^0$
meson is given by

\begin{equation}
	\frac{dN_\gamma}{dE_\gamma} =  \int f_\text{p} (E_\text{p}) f_{ph}(\epsilon) \phi_\gamma(\eta, \text{x}) \frac{dE_{\text{p}}}{dE_{\text{p}}} d\epsilon
\end{equation}

where,{\[f_p(E_p) \, dE_p\] and \[n_{\text{ph}}(\epsilon) \,d\epsilon\]}  are the proton and photon number densities in the energy ranges \(dE_p\) and \(d\epsilon\) respectively.

\begin{figure*}
\begin{center}
\includegraphics[scale=0.5]{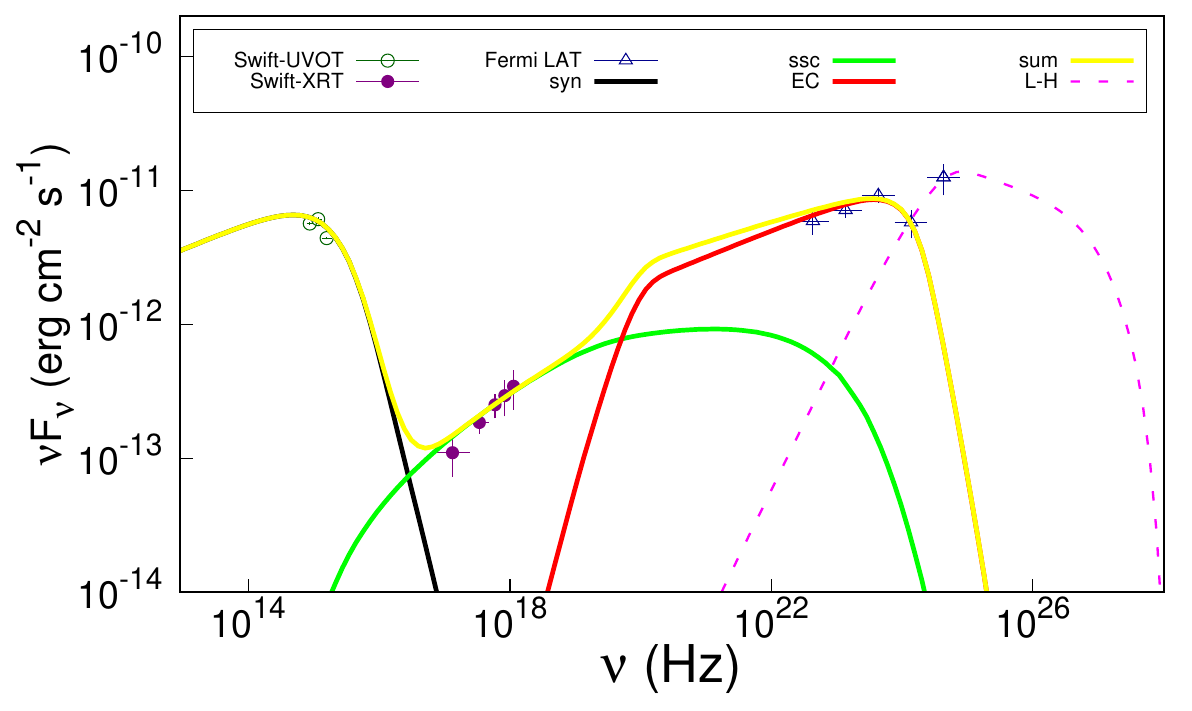}
\caption{Broad band spectral energy distribution along with the one zone leptonic emission model fits for $E_{1}$ with the inclusion of the photo-meson process. The synchrotron emission is represented by the black solid line, the EC emission by the red line, the SSC emission by the green line, and the sum is represented by the yellow solid line. The photo-meson process is shown by the pink dotted line. }
\label{figure-photo}
\end{center}
\end{figure*}

The epoch $E_{1}$ is re-modeled by including a photo-meson process and shown in Fig.  \ref{figure-photo}.

\begin{table}
	\centering
	\caption {Results of the broadband SED analysis carried out for the
epochs $E_{1}$, $E_{2}$ and $E_{3}$. Here, $E_{1}$ represents the quiescent state, $E_{2}$, represents the
active state and
$E_{3}$ represents the high state of the source.  The parameters $p$ and $q$ are the low
and high energy power law indices of the electron distribution, $\gamma_b$ is the the break energy, R is the size of the emission blob in cm, $\Gamma$ is the bulk Lorentz factor and $B$ is the magnetic field in Gauss.}
	\label{table-4}
	\begin{tabular}{lccr} 
		\hline
		Parameter & $E_{1}$  & $E_{2}$  & $E_{3}$ \\
            
		\hline
		p &  2.60 & 2.70 & 2.58\\
		q &  7.0 & 4.40 & 5.57\\
		$\gamma_b$ &7063 & 8067 & 12339\\
        log (R) (cm) &  16.4 & 16.2 & 16.4\\
		$\Gamma$ &  11.3 & 23.1 & 11.2\\
		B (G) &  0.65 & 0.68 & 0.74\\
		\hline
	\end{tabular}
\end{table}

\section{Summary}
\label{sec:dis}
In this work, we carried out $\gamma$-ray spectral, timing, and multi-band SED analysis of 
the neutrino blazar PKS 0735+178 based on $\gamma$-ray data collected over a period of 14 years.
The main motivation is to see how the broadband SED of the blazar during the neutrino
detection epoch compares with the SED of the same source generated at other epochs. The results
are summarised below

\begin{itemize}
\item The source was in its historic brightness state in the $\gamma$-ray band during $E_{3}$.
During this period, a neutrino event was localized close to the vicinity of PKS 0735+178.
During this period, in addition to its brightness state in the $\gamma$-ray band, it also 
flared in the UV, optical and X-ray bands
\item From variability analysis of the three epochs, we found a maximum F$_{var}$ of 95\% during the epoch
$E_{3}$.
\item On shorter time scales, we found a flux doubling/halving time scale of  5.75 hr, and the size of the  $\gamma$-ray emission region is estimated to be 8.22$\times$10$^{15}$ cm.
\item The $\gamma$-ray spectra during all three epochs are well fit with the PL model.
The photon indices obtained by the PL model fit to all three epochs range between 1.9 and 2.
This is similar to that known BL Lacs \citep{2015MNRAS.454..115S}, however, harder when compared to FSRQs.
\item We found the source to show spectral variations, A moderately softer when brighter trend was noticed in all three epochs.
\item The SED modelling of the epochs $E_{2}$ and $E_{3}$ leads to the conclusion that the observed
$\gamma$-ray emission is due to IC scattering of thermal IR photons by relativistic jet
electrons. Even during the epoch of neutrino detection, the SED of PKS 0735+178 can be
well modeled under pure leptonic emission scenario. Probably during this epoch the hadronic emission 
contribution is subdominant to the leptonic process. 
This is consistent with \cite{2023MNRAS.519.1396S}, where the authors showed that the SED of PKS 0735+178 during the neutrino detection epoch was leptonic dominated.

\item The leptonic model alone was unable to reproduce the quiescent state $\gamma$-ray spectrum; however,
a combination with the photo-meson process suggests significant improvement in the spectral fit. 
This suggests that the source PKS 0735+178 has a non-negligible hadronic emission contribution during its
quiescent state but during the flaring periods $E_{2}$ and $E_{3}$ leptonic process dominates over the hadronic process.
\end{itemize}
With the identification of the association of neutrinos with more blazars, followed
by multi band data accumulation and SED modelling, it would be possible to constrain
the leptonic and/or hadronic origin of the high energy $\gamma$-ray emission in blazars.

\section*{Acknowledgements}
Athira M Bharathan acknowledges the Department of Science and Technology (DST) for the INSPIRE FELLOWSHIP (IF200255). Also thank the Center for Research, CHRIST (Deemed to be University) for all their support during the course of this work. Special appreciation is extended to Bhoomika Rajput for her dedicated assistance. In this work, we have used the archival $\gamma$-ray data from Fermi Science Support Center (FSSC). We have also used the \emph{Swift}-XRT/UVOT data from the High Energy Astrophysics Science Archive Research Center (HEASARC).

\section*{DATA AVAILABILITY}
All the data used here for analysis is publicly available and the results are incorporated in the paper.


\label{lastpage}

\bibliographystyle{mnras}
\bibliography{main} 

\begin{thebibliography}{}
\makeatletter
\relax
\def\mn@urlcharsother{\let\do\@makeother \do\$\do\&\do\#\do\^\do\_\do\%\do\~}
\def\mn@doi{\begingroup\mn@urlcharsother \@ifnextchar [ {\mn@doi@} {\mn@doi@[]}}
\def\mn@doi@[#1]#2{\def\@tempa{#1}\ifx\@tempa\@empty \href {http://dx.doi.org/#2} {doi:#2}\else \href {http://dx.doi.org/#2} {#1}\fi \endgroup}
\def\mn@eprint#1#2{\mn@eprint@#1:#2::\@nil}
\def\mn@eprint@arXiv#1{\href {http://arxiv.org/abs/#1} {{\tt arXiv:#1}}}
\def\mn@eprint@dblp#1{\href {http://dblp.uni-trier.de/rec/bibtex/#1.xml} {dblp:#1}}
\def\mn@eprint@#1:#2:#3:#4\@nil{\def\@tempa {#1}\def\@tempb {#2}\def\@tempc {#3}\ifx \@tempc \@empty \let \@tempc \@tempb \let \@tempb \@tempa \fi \ifx \@tempb \@empty \def\@tempb {arXiv}\fi \@ifundefined {mn@eprint@\@tempb}{\@tempb:\@tempc}{\expandafter \expandafter \csname mn@eprint@\@tempb\endcsname \expandafter{\@tempc}}}

\bibitem[\protect\citeauthoryear{{Abdo} et~al.,}{{Abdo} et~al.}{2010}]{2010ApJ...716...30A}
{Abdo} A.~A.,  et~al., 2010, \mn@doi [\apj] {10.1088/0004-637X/716/1/30}, \href {https://ui.adsabs.harvard.edu/abs/2010ApJ...716...30A} {716, 30}

\bibitem[\protect\citeauthoryear{{Abdollahi} et~al.,}{{Abdollahi} et~al.}{2020}]{2020ApJS..247...33A}
{Abdollahi} S.,  et~al., 2020, \mn@doi [\apjs] {10.3847/1538-4365/ab6bcb}, \href {https://ui.adsabs.harvard.edu/abs/2020ApJS..247...33A} {247, 33}

\bibitem[\protect\citeauthoryear{{Acharyya} et~al.,}{{Acharyya} et~al.}{2023}]{2023ApJ...954...70A}
{Acharyya} A.,  et~al., 2023, \mn@doi [\apj] {10.3847/1538-4357/ace327}, \href {https://ui.adsabs.harvard.edu/abs/2023ApJ...954...70A} {954, 70}

\bibitem[\protect\citeauthoryear{{Aharonian}}{{Aharonian}}{2000}]{2000NewA....5..377A}
{Aharonian} F.~A.,  2000, \mn@doi [\na] {10.1016/S1384-1076(00)00039-7}, \href {https://ui.adsabs.harvard.edu/abs/2000NewA....5..377A} {5, 377}

\bibitem[\protect\citeauthoryear{{Andruchow}, {Romero}  \& {Cellone}}{{Andruchow} et~al.}{2005}]{2005A&A...442...97A}
{Andruchow} I.,  {Romero} G.~E.,   {Cellone} S.~A.,  2005, \mn@doi [\aap] {10.1051/0004-6361:20053325}, \href {https://ui.adsabs.harvard.edu/abs/2005A&A...442...97A} {442, 97}

\bibitem[\protect\citeauthoryear{{Angel} \& {Stockman}}{{Angel} \& {Stockman}}{1980}]{1980ARA&A..18..321A}
{Angel} J.~R.~P.,  {Stockman} H.~S.,  1980, \mn@doi [\araa] {10.1146/annurev.aa.18.090180.001541}, \href {https://ui.adsabs.harvard.edu/abs/1980ARA&A..18..321A} {18, 321}

\bibitem[\protect\citeauthoryear{{Arnaud}}{{Arnaud}}{1996}]{1996ASPC..101...17A}
{Arnaud} K.~A.,  1996, in {Jacoby} G.~H.,  {Barnes} J.,  eds,  Astronomical Society of the Pacific Conference Series Vol. 101, Astronomical Data Analysis Software and Systems V. p.~17

\bibitem[\protect\citeauthoryear{{Atwood} et~al.,}{{Atwood} et~al.}{2009}]{2009ApJ...697.1071A}
{Atwood} W.~B.,  et~al., 2009, \mn@doi [\apj] {10.1088/0004-637X/697/2/1071}, \href {https://ui.adsabs.harvard.edu/abs/2009ApJ...697.1071A} {697, 1071}

\bibitem[\protect\citeauthoryear{{Begelman} \& {Sikora}}{{Begelman} \& {Sikora}}{1987}]{1987ApJ...322..650B}
{Begelman} M.~C.,  {Sikora} M.,  1987, \mn@doi [\apj] {10.1086/165760}, \href {https://ui.adsabs.harvard.edu/abs/1987ApJ...322..650B} {322, 650}

\bibitem[\protect\citeauthoryear{{Bhatta}}{{Bhatta}}{2022}]{2022Univ....8..513B}
{Bhatta} G.,  2022, \mn@doi [Universe] {10.3390/universe8100513}, \href {https://ui.adsabs.harvard.edu/abs/2022Univ....8..513B} {8, 513}

\bibitem[\protect\citeauthoryear{{Boettcher}, {Fu}, {Govenor}, {King}  \& {Roustazadeh}}{{Boettcher} et~al.}{2022}]{2022arXiv220412242B}
{Boettcher} M.,  {Fu} M.,  {Govenor} T.,  {King} Q.,   {Roustazadeh} P.,  2022, \mn@doi [arXiv e-prints] {10.48550/arXiv.2204.12242}, \href {https://ui.adsabs.harvard.edu/abs/2022arXiv220412242B} {p. arXiv:2204.12242}

\bibitem[\protect\citeauthoryear{{B{\"o}ttcher}, {Reimer}, {Sweeney}  \& {Prakash}}{{B{\"o}ttcher} et~al.}{2013}]{2013ApJ...768...54B}
{B{\"o}ttcher} M.,  {Reimer} A.,  {Sweeney} K.,   {Prakash} A.,  2013, \mn@doi [\apj] {10.1088/0004-637X/768/1/54}, \href {https://ui.adsabs.harvard.edu/abs/2013ApJ...768...54B} {768, 54}

\bibitem[\protect\citeauthoryear{{Buson} et~al.,}{{Buson} et~al.}{2023}]{2023arXiv230511263B}
{Buson} S.,  et~al., 2023, \mn@doi [arXiv e-prints] {10.48550/arXiv.2305.11263}, \href {https://ui.adsabs.harvard.edu/abs/2023arXiv230511263B} {p. arXiv:2305.11263}

\bibitem[\protect\citeauthoryear{{Carswell}, {Strittmatter}, {Williams}, {Kinman}  \& {Serkowski}}{{Carswell} et~al.}{1974}]{1974ApJ...190L.101C}
{Carswell} R.~F.,  {Strittmatter} P.~A.,  {Williams} R.~E.,  {Kinman} T.~D.,   {Serkowski} K.,  1974, \mn@doi [\apjl] {10.1086/181516}, \href {https://ui.adsabs.harvard.edu/abs/1974ApJ...190L.101C} {190, L101}

\bibitem[\protect\citeauthoryear{{Cerruti}}{{Cerruti}}{2020}]{2020Galax...8...72C}
{Cerruti} M.,  2020, \mn@doi [Galaxies] {10.3390/galaxies8040072}, \href {https://ui.adsabs.harvard.edu/abs/2020Galax...8...72C} {8, 72}

\bibitem[\protect\citeauthoryear{{Dzhilkibaev}, {Suvorova}  \& {Baikal-GVD Collaboration}}{{Dzhilkibaev} et~al.}{2021}]{2021ATel15112....1D}
{Dzhilkibaev} Z.~A.,  {Suvorova} O.,   {Baikal-GVD Collaboration} 2021, The Astronomer's Telegram, \href {https://ui.adsabs.harvard.edu/abs/2021ATel15112....1D} {15112, 1}

\bibitem[\protect\citeauthoryear{{Esposito}, {Walter}, {Jean}, {Tramacere}, {T{\"u}rler}, {L{\"a}hteenm{\"a}ki}  \& {Tornikoski}}{{Esposito} et~al.}{2015}]{2015A&A...576A.122E}
{Esposito} V.,  {Walter} R.,  {Jean} P.,  {Tramacere} A.,  {T{\"u}rler} M.,  {L{\"a}hteenm{\"a}ki} A.,   {Tornikoski} M.,  2015, \mn@doi [\aap] {10.1051/0004-6361/201424644}, \href {https://ui.adsabs.harvard.edu/abs/2015A&A...576A.122E} {576, A122}

\bibitem[\protect\citeauthoryear{{Falomo}, {Treves}  \& {Paiano}}{{Falomo} et~al.}{2021}]{2021ATel15132....1F}
{Falomo} R.,  {Treves} A.,   {Paiano} S.,  2021, The Astronomer's Telegram, \href {https://ui.adsabs.harvard.edu/abs/2021ATel15132....1F} {15132, 1}

\bibitem[\protect\citeauthoryear{{Fang}, {Chen}, {Zhang}  \& {Wu}}{{Fang} et~al.}{2022}]{2022ApJ...933..224F}
{Fang} Y.,  {Chen} Q.,  {Zhang} Y.,   {Wu} J.,  2022, \mn@doi [\apj] {10.3847/1538-4357/ac7647}, \href {https://ui.adsabs.harvard.edu/abs/2022ApJ...933..224F} {933, 224}

\bibitem[\protect\citeauthoryear{{Filippini} et~al.,}{{Filippini} et~al.}{2022}]{2022ATel15290....1F}
{Filippini} F.,  et~al., 2022, The Astronomer's Telegram, \href {https://ui.adsabs.harvard.edu/abs/2022ATel15290....1F} {15290, 1}

\bibitem[\protect\citeauthoryear{{Foschini}, {Ghisellini}, {Tavecchio}, {Bonnoli}  \& {Stamerra}}{{Foschini} et~al.}{2011}]{2011A&A...530A..77F}
{Foschini} L.,  {Ghisellini} G.,  {Tavecchio} F.,  {Bonnoli} G.,   {Stamerra} A.,  2011, \mn@doi [\aap] {10.1051/0004-6361/201117064}, \href {https://ui.adsabs.harvard.edu/abs/2011A&A...530A..77F} {530, A77}

\bibitem[\protect\citeauthoryear{{Fossati}, {Maraschi}, {Celotti}, {Comastri}  \& {Ghisellini}}{{Fossati} et~al.}{1998}]{1998MNRAS.299..433F}
{Fossati} G.,  {Maraschi} L.,  {Celotti} A.,  {Comastri} A.,   {Ghisellini} G.,  1998, \mn@doi [\mnras] {10.1046/j.1365-8711.1998.01828.x}, \href {https://ui.adsabs.harvard.edu/abs/1998MNRAS.299..433F} {299, 433}

\bibitem[\protect\citeauthoryear{{Ghisellini}, {Tavecchio}, {Foschini}  \& {Ghirlanda}}{{Ghisellini} et~al.}{2011}]{2011MNRAS.414.2674G}
{Ghisellini} G.,  {Tavecchio} F.,  {Foschini} L.,   {Ghirlanda} G.,  2011, \mn@doi [\mnras] {10.1111/j.1365-2966.2011.18578.x}, \href {https://ui.adsabs.harvard.edu/abs/2011MNRAS.414.2674G} {414, 2674}

\bibitem[\protect\citeauthoryear{{Giommi} \& {Padovani}}{{Giommi} \& {Padovani}}{2021}]{2021Univ....7..492G}
{Giommi} P.,  {Padovani} P.,  2021, \mn@doi [Universe] {10.3390/universe7120492}, \href {https://ui.adsabs.harvard.edu/abs/2021Univ....7..492G} {7, 492}

\bibitem[\protect\citeauthoryear{{Hartman} et~al.,}{{Hartman} et~al.}{1999}]{1999ApJS..123...79H}
{Hartman} R.~C.,  et~al., 1999, \mn@doi [\apjs] {10.1086/313231}, \href {https://ui.adsabs.harvard.edu/abs/1999ApJS..123...79H} {123, 79}

\bibitem[\protect\citeauthoryear{{IceCube Collaboration}}{{IceCube Collaboration}}{2021}]{2021GCN.31191....1I}
{IceCube Collaboration} 2021, GRB Coordinates Network, \href {https://ui.adsabs.harvard.edu/abs/2021GCN.31191....1I} {31191, 1}

\bibitem[\protect\citeauthoryear{{IceCube Collaboration} et~al.,}{{IceCube Collaboration} et~al.}{2018}]{2018Sci...361.1378I}
{IceCube Collaboration} et~al., 2018, \mn@doi [Science] {10.1126/science.aat1378}, \href {https://ui.adsabs.harvard.edu/abs/2018Sci...361.1378I} {361, eaat1378}

\bibitem[\protect\citeauthoryear{{Kelner} \& {Aharonian}}{{Kelner} \& {Aharonian}}{2008}]{2008PhRvD..78c4013K}
{Kelner} S.~R.,  {Aharonian} F.~A.,  2008, \mn@doi [\prd] {10.1103/PhysRevD.78.034013}, \href {https://ui.adsabs.harvard.edu/abs/2008PhRvD..78c4013K} {78, 034013}

\bibitem[\protect\citeauthoryear{{Konigl}}{{Konigl}}{1981}]{1981ApJ...243..700K}
{Konigl} A.,  1981, \mn@doi [\apj] {10.1086/158638}, \href {https://ui.adsabs.harvard.edu/abs/1981ApJ...243..700K} {243, 700}

\bibitem[\protect\citeauthoryear{{Kramarenko}, {Pushkarev}, {Kovalev}, {Lister}, {Hovatta}  \& {Savolainen}}{{Kramarenko} et~al.}{2022}]{2022MNRAS.510..469K}
{Kramarenko} I.~G.,  {Pushkarev} A.~B.,  {Kovalev} Y.~Y.,  {Lister} M.~L.,  {Hovatta} T.,   {Savolainen} T.,  2022, \mn@doi [\mnras] {10.1093/mnras/stab3358}, \href {https://ui.adsabs.harvard.edu/abs/2022MNRAS.510..469K} {510, 469}

\bibitem[\protect\citeauthoryear{{Lynden-Bell}}{{Lynden-Bell}}{1969}]{1969Natur.223..690L}
{Lynden-Bell} D.,  1969, \mn@doi [\nat] {10.1038/223690a0}, \href {https://ui.adsabs.harvard.edu/abs/1969Natur.223..690L} {223, 690}

\bibitem[\protect\citeauthoryear{{Mannheim}}{{Mannheim}}{1993a}]{1993PhRvD..48.2408M}
{Mannheim} K.,  1993a, \mn@doi [\prd] {10.1103/PhysRevD.48.2408}, \href {https://ui.adsabs.harvard.edu/abs/1993PhRvD..48.2408M} {48, 2408}

\bibitem[\protect\citeauthoryear{{Mannheim}}{{Mannheim}}{1993b}]{1993A&A...269...67M}
{Mannheim} K.,  1993b, \mn@doi [\aap] {10.48550/arXiv.astro-ph/9302006}, \href {https://ui.adsabs.harvard.edu/abs/1993A&A...269...67M} {269, 67}

\bibitem[\protect\citeauthoryear{{Mattox} et~al.,}{{Mattox} et~al.}{1996}]{1996ApJ...461..396M}
{Mattox} J.~R.,  et~al., 1996, \mn@doi [\apj] {10.1086/177068}, \href {https://ui.adsabs.harvard.edu/abs/1996ApJ...461..396M} {461, 396}

\bibitem[\protect\citeauthoryear{{Max-Moerbeck} et~al.,}{{Max-Moerbeck} et~al.}{2014}]{2014MNRAS.445..428M}
{Max-Moerbeck} W.,  et~al., 2014, \mn@doi [\mnras] {10.1093/mnras/stu1749}, \href {https://ui.adsabs.harvard.edu/abs/2014MNRAS.445..428M} {445, 428}

\bibitem[\protect\citeauthoryear{{Nolan} et~al.,}{{Nolan} et~al.}{2012}]{2012ApJS..199...31N}
{Nolan} P.~L.,  et~al., 2012, \mn@doi [\apjs] {10.1088/0067-0049/199/2/31}, \href {https://ui.adsabs.harvard.edu/abs/2012ApJS..199...31N} {199, 31}

\bibitem[\protect\citeauthoryear{Padovani, Oikonomou, Petropoulou, Giommi  \& Resconi}{Padovani et~al.}{2019}]{10.1093/mnrasl/slz011}
Padovani P.,  Oikonomou F.,  Petropoulou M.,  Giommi P.,   Resconi E.,  2019, \mn@doi [Monthly Notices of the Royal Astronomical Society: Letters] {10.1093/mnrasl/slz011}, 484, L104

\bibitem[\protect\citeauthoryear{{Paliya}, {B{\"o}ttcher}, {Olmo-Garc{\'\i}a}, {Dom{\'\i}nguez}, {Gil de Paz}, {Franckowiak}, {Garrappa}  \& {Stein}}{{Paliya} et~al.}{2020}]{2020ApJ...902...29P}
{Paliya} V.~S.,  {B{\"o}ttcher} M.,  {Olmo-Garc{\'\i}a} A.,  {Dom{\'\i}nguez} A.,  {Gil de Paz} A.,  {Franckowiak} A.,  {Garrappa} S.,   {Stein} R.,  2020, \mn@doi [\apj] {10.3847/1538-4357/abb46e}, \href {https://ui.adsabs.harvard.edu/abs/2020ApJ...902...29P} {902, 29}

\bibitem[\protect\citeauthoryear{{Pandey} \& {Stalin}}{{Pandey} \& {Stalin}}{2022}]{2022A&A...668A.152P}
{Pandey} A.,  {Stalin} C.~S.,  2022, \mn@doi [\aap] {10.1051/0004-6361/202244648}, \href {https://ui.adsabs.harvard.edu/abs/2022A&A...668A.152P} {668, A152}

\bibitem[\protect\citeauthoryear{{Pandey}, {Rajput}  \& {Stalin}}{{Pandey} et~al.}{2022}]{2022MNRAS.510.1809P}
{Pandey} A.,  {Rajput} B.,   {Stalin} C.~S.,  2022, \mn@doi [\mnras] {10.1093/mnras/stab3338}, \href {https://ui.adsabs.harvard.edu/abs/2022MNRAS.510.1809P} {510, 1809}

\bibitem[\protect\citeauthoryear{{Petkov}, {Novoseltsev}, {Novoseltseva}  \& {Baksan Underground Scintillation Telescope Group}}{{Petkov} et~al.}{2021}]{2021ATel15143....1P}
{Petkov} V.~B.,  {Novoseltsev} Y.~F.,  {Novoseltseva} R.~V.,   {Baksan Underground Scintillation Telescope Group} 2021, The Astronomer's Telegram, \href {https://ui.adsabs.harvard.edu/abs/2021ATel15143....1P} {15143, 1}

\bibitem[\protect\citeauthoryear{{Plavin}, {Kovalev}, {Kovalev}  \& {Troitsky}}{{Plavin} et~al.}{2020}]{2020ATel14238....1P}
{Plavin} A.~V.,  {Kovalev} Y.~Y.,  {Kovalev} Y.~A.,   {Troitsky} S.~V.,  2020, The Astronomer's Telegram, \href {https://ui.adsabs.harvard.edu/abs/2020ATel14238....1P} {14238, 1}

\bibitem[\protect\citeauthoryear{{Plavin}, {Burenin}, {Kovalev}, {Lutovinov}, {Starobinsky}, {Troitsky}  \& {Zakharov}}{{Plavin} et~al.}{2023a}]{2023arXiv230600960P}
{Plavin} A.~V.,  {Burenin} R.~A.,  {Kovalev} Y.~Y.,  {Lutovinov} A.~A.,  {Starobinsky} A.~A.,  {Troitsky} S.~V.,   {Zakharov} E.~I.,  2023a, \mn@doi [arXiv e-prints] {10.48550/arXiv.2306.00960}, \href {https://ui.adsabs.harvard.edu/abs/2023arXiv230600960P} {p. arXiv:2306.00960}

\bibitem[\protect\citeauthoryear{{Plavin}, {Kovalev}, {Kovalev}  \& {Troitsky}}{{Plavin} et~al.}{2023b}]{2023MNRAS.523.1799P}
{Plavin} A.~V.,  {Kovalev} Y.~Y.,  {Kovalev} Y.~A.,   {Troitsky} S.~V.,  2023b, \mn@doi [\mnras] {10.1093/mnras/stad1467}, \href {https://ui.adsabs.harvard.edu/abs/2023MNRAS.523.1799P} {523, 1799}

\bibitem[\protect\citeauthoryear{{Poole} et~al.,}{{Poole} et~al.}{2008}]{2008MNRAS.383..627P}
{Poole} T.~S.,  et~al., 2008, \mn@doi [\mnras] {10.1111/j.1365-2966.2007.12563.x}, \href {https://ui.adsabs.harvard.edu/abs/2008MNRAS.383..627P} {383, 627}

\bibitem[\protect\citeauthoryear{{Prince}, {Das}, {Gupta}, {Majumdar}  \& {Czerny}}{{Prince} et~al.}{2023}]{2023arXiv230106565P}
{Prince} R.,  {Das} S.,  {Gupta} N.,  {Majumdar} P.,   {Czerny} B.,  2023, \mn@doi [arXiv e-prints] {10.48550/arXiv.2301.06565}, \href {https://ui.adsabs.harvard.edu/abs/2023arXiv230106565P} {p. arXiv:2301.06565}

\bibitem[\protect\citeauthoryear{{Rajput}, {Stalin}, {Sahayanathan}, {Rakshit}  \& {Mandal}}{{Rajput} et~al.}{2019}]{2019MNRAS.486.1781R}
{Rajput} B.,  {Stalin} C.~S.,  {Sahayanathan} S.,  {Rakshit} S.,   {Mandal} A.~K.,  2019, \mn@doi [\mnras] {10.1093/mnras/stz941}, \href {https://ui.adsabs.harvard.edu/abs/2019MNRAS.486.1781R} {486, 1781}

\bibitem[\protect\citeauthoryear{{Rajput}, {Stalin}  \& {Sahayanathan}}{{Rajput} et~al.}{2020}]{2020MNRAS.498.5128R}
{Rajput} B.,  {Stalin} C.~S.,   {Sahayanathan} S.,  2020, \mn@doi [\mnras] {10.1093/mnras/staa2708}, \href {https://ui.adsabs.harvard.edu/abs/2020MNRAS.498.5128R} {498, 5128}

\bibitem[\protect\citeauthoryear{{Rajput}, {Shah}, {Stalin}, {Sahayanathan}  \& {Rakshit}}{{Rajput} et~al.}{2021}]{2021MNRAS.504.1772R}
{Rajput} B.,  {Shah} Z.,  {Stalin} C.~S.,  {Sahayanathan} S.,   {Rakshit} S.,  2021, \mn@doi [\mnras] {10.1093/mnras/stab970}, \href {https://ui.adsabs.harvard.edu/abs/2021MNRAS.504.1772R} {504, 1772}

\bibitem[\protect\citeauthoryear{{Rakshit}, {Stalin}, {Muneer}, {Neha}  \& {Paliya}}{{Rakshit} et~al.}{2017}]{2017ApJ...835..275R}
{Rakshit} S.,  {Stalin} C.~S.,  {Muneer} S.,  {Neha} S.,   {Paliya} V.~S.,  2017, \mn@doi [\apj] {10.3847/1538-4357/835/2/275}, \href {https://ui.adsabs.harvard.edu/abs/2017ApJ...835..275R} {835, 275}

\bibitem[\protect\citeauthoryear{{Rani}, {Krichbaum}, {Marscher}, {Jorstad}, {Hodgson}, {Fuhrmann}  \& {Zensus}}{{Rani} et~al.}{2014}]{2014A&A...571L...2R}
{Rani} B.,  {Krichbaum} T.~P.,  {Marscher} A.~P.,  {Jorstad} S.~G.,  {Hodgson} J.~A.,  {Fuhrmann} L.,   {Zensus} J.~A.,  2014, \mn@doi [\aap] {10.1051/0004-6361/201424796}, \href {https://ui.adsabs.harvard.edu/abs/2014A&A...571L...2R} {571, L2}

\bibitem[\protect\citeauthoryear{{Rector} \& {Stocke}}{{Rector} \& {Stocke}}{2001}]{2001AJ....122..565R}
{Rector} T.~A.,  {Stocke} J.~T.,  2001, \mn@doi [\aj] {10.1086/321179}, \href {https://ui.adsabs.harvard.edu/abs/2001AJ....122..565R} {122, 565}

\bibitem[\protect\citeauthoryear{{Sahakyan}, {Giommi}, {Padovani}, {Petropoulou}, {B{\'e}gu{\'e}}, {Boccardi}  \& {Gasparyan}}{{Sahakyan} et~al.}{2023}]{2023MNRAS.519.1396S}
{Sahakyan} N.,  {Giommi} P.,  {Padovani} P.,  {Petropoulou} M.,  {B{\'e}gu{\'e}} D.,  {Boccardi} B.,   {Gasparyan} S.,  2023, \mn@doi [\mnras] {10.1093/mnras/stac3607}, \href {https://ui.adsabs.harvard.edu/abs/2023MNRAS.519.1396S} {519, 1396}

\bibitem[\protect\citeauthoryear{{Sahayanathan}, {Sinha}  \& {Misra}}{{Sahayanathan} et~al.}{2018}]{2018RAA....18...35S}
{Sahayanathan} S.,  {Sinha} A.,   {Misra} R.,  2018, \mn@doi [Research in Astronomy and Astrophysics] {10.1088/1674-4527/18/3/35}, \href {https://ui.adsabs.harvard.edu/abs/2018RAA....18...35S} {18, 035}

\bibitem[\protect\citeauthoryear{{Saito}, {Stawarz}, {Tanaka}, {Takahashi}, {Madejski}  \& {D'Ammando}}{{Saito} et~al.}{2013}]{2013ApJ...766L..11S}
{Saito} S.,  {Stawarz} {\L}.,  {Tanaka} Y.~T.,  {Takahashi} T.,  {Madejski} G.,   {D'Ammando} F.,  2013, \mn@doi [\apjl] {10.1088/2041-8205/766/1/L11}, \href {https://ui.adsabs.harvard.edu/abs/2013ApJ...766L..11S} {766, L11}

\bibitem[\protect\citeauthoryear{{Shakura} \& {Sunyaev}}{{Shakura} \& {Sunyaev}}{1973}]{1973A&A....24..337S}
{Shakura} N.~I.,  {Sunyaev} R.~A.,  1973, \aap, \href {https://ui.adsabs.harvard.edu/abs/1973A&A....24..337S} {24, 337}

\bibitem[\protect\citeauthoryear{{Singal}}{{Singal}}{2015}]{2015MNRAS.454..115S}
{Singal} J.,  2015, \mn@doi [\mnras] {10.1093/mnras/stv1964}, \href {https://ui.adsabs.harvard.edu/abs/2015MNRAS.454..115S} {454, 115}

\bibitem[\protect\citeauthoryear{{Ulrich}, {Maraschi}  \& {Urry}}{{Ulrich} et~al.}{1997}]{1997ARA&A..35..445U}
{Ulrich} M.-H.,  {Maraschi} L.,   {Urry} C.~M.,  1997, \mn@doi [\araa] {10.1146/annurev.astro.35.1.445}, \href {https://ui.adsabs.harvard.edu/abs/1997ARA&A..35..445U} {35, 445}

\bibitem[\protect\citeauthoryear{{Vaughan}, {Edelson}, {Warwick}  \& {Uttley}}{{Vaughan} et~al.}{2003}]{2003MNRAS.345.1271V}
{Vaughan} S.,  {Edelson} R.,  {Warwick} R.~S.,   {Uttley} P.,  2003, \mn@doi [\mnras] {10.1046/j.1365-2966.2003.07042.x}, \href {https://ui.adsabs.harvard.edu/abs/2003MNRAS.345.1271V} {345, 1271}

\bibitem[\protect\citeauthoryear{{Wagner} \& {Witzel}}{{Wagner} \& {Witzel}}{1995}]{1995ARA&A..33..163W}
{Wagner} S.~J.,  {Witzel} A.,  1995, \mn@doi [\araa] {10.1146/annurev.aa.33.090195.001115}, \href {https://ui.adsabs.harvard.edu/abs/1995ARA&A..33..163W} {33, 163}

\bibitem[\protect\citeauthoryear{{Yuan} et~al.,}{{Yuan} et~al.}{2023}]{2023ApJ...953...47Y}
{Yuan} Q.,  et~al., 2023, \mn@doi [\apj] {10.3847/1538-4357/acdd74}, \href {https://ui.adsabs.harvard.edu/abs/2023ApJ...953...47Y} {953, 47}

\makeatother
\end{thebibliography}
\end{document}